\documentclass[aps,prc,preprint,superscriptaddress,showpacs,amssymb,amsmath,amsfonts,floatfix]{revtex4}
\usepackage{graphicx}
\newcommand{\be}{\begin{equation}}
\newcommand{\ee}{\end{equation}}
\newcommand{\beq}{\begin{eqnarray}}
\newcommand{\eeq}{\end{eqnarray}}
\newcommand{\ba}{\begin{array}}
\newcommand{\ea}{\end{array}}
\newcommand{\gn}{\gamma N\!\to\!\pi N}
\newcommand{\gpp}{\gamma p\!\to\!\pi^0 p}

\newcommand{\gnp}{\gamma n\!\to\!\pi^- p}

\newcommand{\gd}{\gamma d\!\to\!\pi NN}

\newcommand{\gdpp}{\gamma d\!\to\!\pi^- pp}
\newcommand{\LRA}{\!\leftrightarrow\!}

\newcommand{\bgnp}{\mbox {\boldmath $\gamma n\!\to\!\pi^-p\,$}}

\newcommand{\bgdpp}{\mbox {\boldmath $\gamma d\to\pi^- pp$}}
\newcommand{\mgn}{M_{\gamma N}}
\newcommand{\hmgn}{\hat M_{\gamma N}}
\newcommand{\mgne}{M_{\gamma n}}
\newcommand{\hmgne}{\hat M_{\gamma n}}
\newcommand{\mgd}{M_{\gamma d}}
\newcommand{\hmgd}{\hat M_{\gamma d}}
\newcommand{\hgn}{\hat G_N}
\newcommand{\hgd}{\hat\Gamma_d}
\newcommand{\bfq}{\mbox {\boldmath $q$}}
\newcommand{\bfk}{\mbox {\boldmath $k$}}
\newcommand{\bfkk}{\mbox {\boldmath $K$}}
\newcommand{\bfss}{\mbox {\boldmath $S$}}
\newcommand{\bfvv}{\mbox {\boldmath $V$}}
\newcommand{\bfp}{\mbox {\boldmath $p$}}
\newcommand{\bfr}{\mbox {\boldmath $r$}}

\newcommand{\bfx}{{\bf x}}
\newcommand{\bpN}{\bfp^{}_N}
\newcommand{\pN}{p^{}_N}
\newcommand{\bpNi}{\bfp^{\,\prime}_N}
\newcommand{\pNi}{p^{\,\prime}_N}
\newcommand{\bki}{\bfk^{\,\prime}_2}
\newcommand{\ki}{k^{\,\prime}_2}
\newcommand{\sss}{\!\!\!/}
\newcommand{\bfsig}{\mbox {\boldmath $\sigma$}}
\newcommand{\bfe}{\mbox {\boldmath $e$}}

\newcommand{\bfn}{\mbox {\boldmath $n^{}$}}
\newcommand{\bfm}{\mbox {\boldmath $m$}}
\newcommand{\bfl}{\mbox {\boldmath $l$}}
\newcommand{\bfa}{\mbox {\boldmath $a$}}
\newcommand{\bfb}{\mbox {\boldmath $b$}}
\newcommand{\hff}{\hat F}
\newcommand{\hqq}{\hat Q}
\newcommand{\half}{\frac{1}{2}}
\newcommand{\dif}{differential}
\newcommand{\crs}{cross section}
\newcommand{\crss}{\crs s}
\newcommand{\ega}{E_{\gamma}}
\newcommand{\dsdo}{\frac {d\sigma}{d\Omega}}
\newcommand{\dso}{d\sigma\!/d\Omega}
\newcommand{\sgd}{\sigma_{\gamma d}}
\newcommand{\sgn}{\sigma_{\gamma n}}
\newcommand{\asgn}{\bar{\sigma}_{\gamma n}}
\newcommand{\dist}{\displaystyle}
\newcommand{\hmm}{\hat M}

\newcommand{\he}{\hat e}

\newcommand{\hn}{\hat n}

\newcommand{\huu}{\hat U}

\newcommand{\rtt}{\rule{0pt}{12pt}}
\newcommand{\rpt}{\rule{0pt}{14pt}}
\newcommand{\rptt}{\rule{0pt}{18pt}}
\newcommand{\rppt}{\rule{0pt}{22pt}}
\newcommand{\rpptt}{\rule{0pt}{24pt}}

\newcommand{\hPsi}{\hat\Psi_d}
\newcommand{\bfeps}{\mbox {\boldmath $\epsilon$}}

\newcommand{\ointpi}{\oint\frac{d\bpNi}{(2\pi)^3}}
\newcommand{\intdpi}{\int\!\!\frac{d\bpNi}{(2\pi)^3}}
\newcommand{\intdpp}{\int\!\!\frac{d^{4\!}p\,'_2}{(2\pi)^4}\,}
\newcommand{\intdki}{\int\!\!\frac{d\bki}{(2\pi)^3}}
\newcommand{\ointp}{\oint\frac{d\pNi\,p^{\,\prime\,2}_N}{\pi E'}}
\newcommand{\ointk}{\oint\frac{d\ki\,k^{\,\prime\,2}_2}{\pi E'}}
\newcommand{\abra}{\left(\!\! \begin{array}{cc} }
\newcommand{\aket}{\end{array}\!\!\right)}

\newcommand{\tff}{\tilde F}

\begin{document}

\title{Extracting the photoproduction \crs\ off the neutron
       $\bgnp$ from deuteron data with FSI effects}

\newcommand*{\GWU}{Center for Nuclear Studies, Department of Physics, \\
            The George Washington University, Washington, DC 20052, USA}
\newcommand*{\Duke}{Duke University, Durham, NC 27708, USA}
\newcommand*{\ITEP}{Institute of Theoretical and Experimental Physics,
            Moscow, 117259 Russia}
\author {V.E.~Tarasov}
\affiliation{\ITEP}
\author {W.J.~Briscoe}
\affiliation{\GWU}
\author {H.~Gao}
\affiliation{\Duke}
\author {A.E.~Kudryavtsev}
\affiliation{\ITEP}
\affiliation{\GWU}
\author {I.I.~Strakovsky}
\affiliation{\GWU}

\begin{abstract}
The incoherent pion photoproduction reaction $\gdpp$ is considered theoretically 
in a wide energy region $E_{th}\le\ega\le 2700$~MeV. The model applied contains 
the impulse approximation as well as the NN- and $\pi N$-FSI amplitudes. The aim 
of the paper is to study a reliable way for getting the information on elementary 
$\gnp$ reaction \crs\ beyond the impulse approximation for $\gdpp$. For the 
elementary $\gamma N\!\to\!\pi N$, $N\!N\!\to\!N\!N$, and $\pi N\!\to\!\pi N$ 
amplitudes, the results of the GW DAC are used. There are no additional theoretical 
constraints. The calculated \crss\ $\dso(\gdpp)$ are compared with existing data.
The procedure used to extract information on the \dif\ \crs\ $\dso(\gnp)$ on the 
neutron from the deuteron data using the FSI correction factor $R$ is discussed. 
The calculations for $R$ versus $\pi^-p$ CM angle $\theta_1$ of the outgoing pion 
are performed at different photon-beam energies with kinematical cuts for 
``quasi-free" process $\gnp$. The results show a sizeable FSI effect $R\ne 1$ from 
$S$-wave part of $pp$-FSI at small angles close to $\theta_1\sim 0$: this region 
narrows as the photon energy increases. At larger angles, the effect is small 
($|R\!-\!1|\ll 1$) and agrees with estimations of FSI in the Glauber approach.
\end{abstract}

\pacs{13.60.Le, 21.45.Bc, 24.10.Eq, 25.20.Lj}
\maketitle

\section{Introduction}
\label{sec:intro}

The $N^\ast$ family of nucleon resonances has many well-established members~\cite{PDG}, 
several of which exhibit overlapping resonances with very similar masses and widths, 
but with different $J^P$ spin-parity values.  Apart from the $N(1535)1/2^-$ state, the 
known proton and neutron photo-decay amplitudes have been determined from analyses of 
single-pion photoproduction. The present work studies the region from threshold to the 
upper limit of the SAID analyses, which is $W=2.5$~GeV.  There are two closely spaced 
states above the $\Delta(1232)3/2^+$: $N(1520)3/2^-$ and $N(1535)1/2^-$. Up to a CM 
energy of $W\approx 1800$~MeV, this region also encompasses a sequence of six overlapping 
states: $N(1650)1/2^-$, $N(1675)5/2^-$, $N(1680)5/2^+$, $N(1700)3/2^-$, $N(1710)1/2^+$, 
and $N(1720)3/2^+$.

One critical issue in the study of meson photoproduction on the nucleon comes from isospin. 
While isospin can change at the photon vertex, it must be conserved at the final hadronic 
vertex. Only with good data on both proton and neutron targets can one  hope to disentangle 
the isoscalar and isovector electromagnetic couplings of the various N$^\ast$ and $\Delta^\ast$ 
resonances (see, Refs.~\cite{Wat,Walker}), as well as the isospin properties of the non-resonant 
background amplitudes. The lack of $\gamma n\to\pi^-p$ and $\pi^0n$ data does not allow us
to be as confident about the determination of neutron couplings relative to those of the proton.
Some of the $N^{\ast}$ baryons ($N(1675)5/2^-$, for instance) have stronger electromagnetic 
couplings to the neutron relative to the proton, but the parameters are very uncertain~\cite{PDG}.
Data on the $\gn$ reactions are needed to improve the amplitudes and expand them to higher energies.

Incoherent pion photoproduction on the deuteron is interesting in various aspects of nuclear 
physics, and particularly provides information on the elementary reaction on the neutron, i.e., 
$\gamma n\!\to\!\pi N$. Final-state-interaction (FSI) plays a critical role in the state-of-the-art 
analysis of the $\gn$ interaction as extracted from $\gd$ data. The FSI was first considered in 
Refs.~\cite{Migdal,Watson} as responsible for the near threshold enhancement (Migdal-Watson effect) 
in the $NN$-mass spectrum of the meson production reaction $NN\to NNx$. In Ref.~\cite{Baru}, the FSI 
amplitude was studied in detail. Calculations of $NN$- and $\pi N$-FSI for the reactions $\gd$ can 
be traced back to Refs.~\cite{BL,La78,La81}. In Refs.~\cite{La78,La81}, the elementary $\gn$ amplitude, 
constructed in Ref.~\cite{BL} from the Born terms and $\Delta(1232)3/2^+$ contribution, was used in 
$\gd$ calculations with FSI terms taken into account. Good descriptions of the available deuteron 
data for charged pion photoproduction in the threshold and $\Delta(1232)3/2^+$ regions were obtained.

Further developments of this topic (see \cite{Dar,DAS,Fix,Lev06,Schwamb,Lev10, Dar1,La06} and 
references therein) included improvements of the elementary $\gn$ amplitude, predictions for the 
unpolarized and polarized (polarized beam, target or both, see~\cite{Dar,Fix,Lev06,Schwamb,Lev10,Dar1} 
and references therein) observables in the $\gd$ reactions, and comparison with new data. Different 
models for $\gn$ amplitude were used in the above mentioned papers, \emph{i.e.}, MAID~\cite{MAID98} 
(Refs.~\cite{Fix,Lev06}), SAID~\cite{SAID02} (Refs.~\cite{Lev06,Lev10}), and MAID~\cite{MAID07} 
(Ref.~\cite{Lev10}). As discussed in Refs.~\cite{Lev06,Lev10}, the main uncertainties of $\gd$ 
calculations stem from the model dependence of the $\gn$ amplitude. In the latest SAID~\cite{SAID02} 
and MAID~\cite{MAID07} analyses, the models for $\gn$ amplitudes are developed for the photon energies 
$\ega<2.7$~GeV~\cite{SAID02} and $\ega<1.65$~GeV~\cite{MAID07}, respectively. Summary results from the 
existing $\gd$ calculations show that FSI effects significantly reduce the \dif\ \crs\ for $\pi^0 pn$ 
channel, mainly due to the $pn$ rescattering, and contribute much less in the charged-pion case, i.e., 
in $\pi^+nn$ and $\pi^-pp$ channels.

The role of FSI depends on the kinematical region considered. In Ref.~\cite{KDT}, a narrow enhancement 
in the $pp$-mass spectrum observed in the reaction $pp\!\to\!pp\pi^-$ with backward outgoing $\pi^-$ 
was explained by the $pp$-FSI. The result was shown to be model-independent, determined only by 
$pp$-scattering parameters for the $pp$ pair produced at high momentum transfer. In the same approach, 
it was shown~\cite{DKT} that the observed energy behavior of the total \crs\ of the reaction 
$pp\!\to\!pp\eta$ in the near threshold region can be also explained by $pp$-FSI. In Ref.~\cite{La06}, 
the meson photoproducton on deuteron was considered at high energies ($\ega\sim$~several GeV) and high 
momentum transferred to final meson. This work was focused mainly on special kinematical regions close 
to the logarithmic singularities of the triangle NN- and $\pi$N-FSI amplitudes, the latter are strongly 
enhanced. These configurations where the FSI amplitudes dominates may be interesting, say, in connection 
with color transparency hypothesis~\cite{CT}. On the other hand, to extract the neutron data, we are 
interested in the opposite case, i.e., when FSI is suppressed.

In this paper, the role of FSI in the $\gdpp$ reaction is under consideration. Our analysis addresses 
the data~\cite{Wei,CLAS} that come from the $\gdpp$ experiment at JLab using CLAS for a wide range of 
photon-beam energies up to about 3.5~GeV. The calculated FSI corrections for this reaction are further 
used to extract the $\gnp$ data that constrain the $\gamma N\to\pi N$ amplitude used in PWA and coupled 
channel technologies.

\begin{figure}
\begin{center}
\begin{minipage}[b]{15cm}
\includegraphics[width=4.4cm, keepaspectratio]{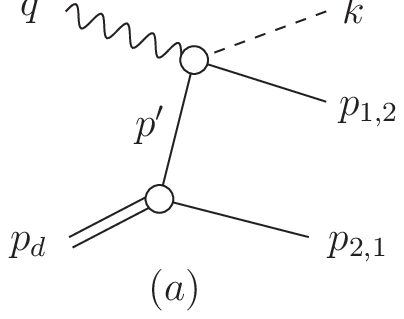}~~~~~
\includegraphics[width=4.4cm, keepaspectratio]{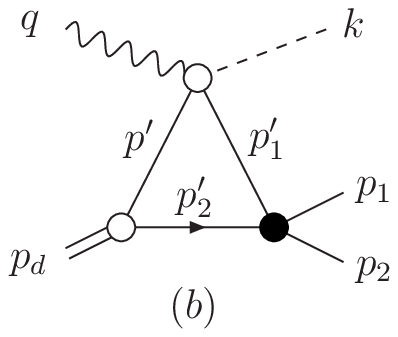}~~~~~
\includegraphics[width=4.4cm, keepaspectratio]{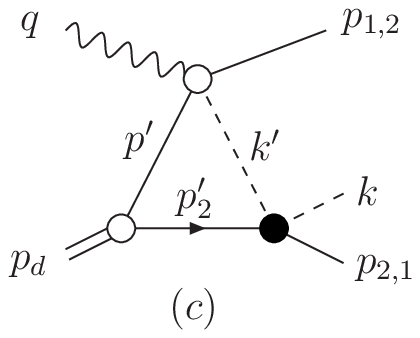}
\end{minipage}
\end{center}
\caption{Feynman diagrams for the leading components of the
        $\gamma d\to\pi^-pp$ amplitude. (a) Impulse approximation,
        (b) $pp$-FSI, and (c) $\pi$N-FSI. Filled black circles
        show FSI vertices. Wavy, dashed, solid, and double lines
        correspond to the photons, pions, nucleons, and deuterons,
        respectively.} \label{fig:g1}
\end{figure}

In our approach, the $\gdpp$ amplitude has three leading terms, represented by the diagrams in 
Fig.~\ref{fig:g1}: impulse approximation (IA) [Fig.~\ref{fig:g1}(a)], $pp$-FSI [Fig.~\ref{fig:g1}(b)], 
and $\pi$N-FSI [Fig.~\ref{fig:g1}(c)] contributions. IA and $\pi$N diagrams [Figs.~\ref{fig:g1}(a),(c)] 
include also the cross-terms between outgoing protons. It is convenient to study the FSI effects in 
terms of the ratio
\be
	R^{}_{FSI}=(d\sigma/d\Omega_{\pi p})/(d\sigma^{IA}/d\Omega_{\pi p}),
\label{i1}\ee
\emph{i.e.}, the ratio of the \dif\ \crss\ $\dso_{\pi p}$ including the full calculations of diagrams 
[Figs.~\ref{fig:g1}(a)--(c)] to the ($d\sigma^{IA}/d\Omega_{\pi p}$), associated with IA diagram 
[Fig.~\ref{fig:g1}(a)], where $\Omega_{\pi p}$ is the solid angle of the relative motion in the final 
$\pi p$ system. The ratio $R^{}_{FSI}$~(\ref{i1}) depends on different kinematical variables.  It can 
be used to extract the \dif\ \crss\ $\dso$ for the reaction $\gnp$ from the $\gdpp$ data. We use the 
recent GW pion photoproduction multipoles to constrain the amplitude for the impulse 
approximation~\cite{pr_PWA} with no additional theoretical input. While for the $pp$-FSI and $\pi$N-FSI, 
we include the GW $N\!N$~\cite{NN_PWA} and GW $\pi$N amplitudes~\cite{piN_PWA}, respectively, for the 
deuteron description, we use the wave function of the CD-Bonn potential~\cite{BonnCD} with S- and D-wave
components included.

This paper is organized as follows. In Section~\ref{sec:model}, we
describe the model. In Subsections~\ref{sec:kin} and \ref{sec:ImpApp},
we introduce the notations and write out the impulse approximation
terms of the $\gd$ amplitude. In Subsections~\ref{sec:NN_FSI} and
\ref{sec:piN_FSI}, we derive the $N\!N$-FSI and $\pi\!N$-FSI terms
of the reaction amplitude, respectively.

The results are presented in Section~~\ref{sec:results}.  In
Subsection~\ref{sec:compar}, we compare our numerical results for the
\crs\ $\dso(\gdpp)$ with the DESY data and discuss the contributions
from different amplitudes.  In Subsection~\ref{sec:extract}, we
discuss the procedure to extract the \crs\ $\dso(\gnp)$ for the neutron
from the $\gdpp\,$ data and define the correction factor $R$. In
Subsection~\ref{sec:r-factor}, we present the numerical results for
the factor $R$ and discuss the role of the $S$-wave $pp$-FSI.  
In Subsection~\ref{sec:glauber}, we estimate the $R$ factor in the 
Glauber approach.  The conclusion is given in Section~\ref{sec:conclusion}.

\section{Model for $\bgdpp$ amplitude}
\label{sec:model}

\subsection{Kinematical notations}
\label{sec:kin}

Hereafter $m$, $\mu$, and $m_d$ are the proton, pion, and deuteron
masses, respectively; $q=\!(\ega,\bfq)$, $p_d=\!(E_d,\bfp_d)$,
$k=\!(\omega,\bfk)$, and $p_i=\!(E_i,\bfp_i)$ ($i\!=\!1,2$) are the
4-momenta of the initial photon, deuteron and final pion, nucleons,
respectively; $k'=\!(\omega',\bfk')$, $p'=\!(E',\bfp')$, and $p'_i=\!(E'_i,\bfp'_i)$
are the 4-momenta of the intermediate particles. The 4-momenta are shown in Fig.~\ref{fig:g1}. 
The total energies $\ega,\,E_d,\,\cdots\,E'_i$ and 3-momenta $\bfq,\,\bfp_d,\,\cdots\,\bfp'_i$ 
are given in the laboratory system (LS), \emph{i.e.}, in the deuteron rest frame, where 
$\bfp_d=0$ and $E_d=m_d$.

The cross-section element $d\sigma(\gdpp)$, according to the usual 
conventions for invariant amplitudes and phase spaces (see 
Appendix~\ref{app:amp}), can be written in the form
\be
	d\sigma=\half\,\frac{\overline{|\mgd|^2}}{4\ega m_d}\,d\tau_3,~~~
	d\tau_3=\frac{d\,^3p_2}{(2\pi)^3\,2E_2}\,d\tau_2,~~~
	d\tau_2=\frac{k_1 d\Omega_1}{(4\pi)^2 \ W_1}.
\label{kin1}\ee
Here: $\mgd$ is the $\gdpp$ invariant amplitude; $\overline{|\mgd|^2}$ 
is the square $|\mgd|^2$, calculated for unpolarized particles; $d\tau_3$ 
is the $\pi NN$ phase space element, written in terms of the $\pi p_1$-pair 
phase space element $d\tau_2$ and 3-momentum $\bfp_2$ of the 2nd proton; 
the factor $\half$ in $d\sigma$~(\ref{kin1}) takes into account that the 
final protons are identical; $k_1$ and $\Omega_1$ are the
relative momentum and solid angle of relative motion in the
$\pi p_1$ system, respectively; $W_1$ is the effective mass of the 
$\pi p_1$ system.

\subsection{Impulse-approximation amplitudes}
\label{sec:ImpApp}

Let us use the formalism of Ref.~\cite{Tar}, which is similar to that of 
Gross~\cite{Gross} in the case of small nucleon momenta $|\bfp|^2/m\ll m$ 
in the deuteron vertex. Then, the impulse-approximation term $M_a$ 
[Fig.~\ref{fig:g1}(a)] of the $\gd$ amplitude can be written in the form
\be
	M_a=M^{(1)}_a+M^{(2)}_a,
\label{ma1}\ee
$$
M^{(1)}_a=\bar u_1\,\hmgn^{(1)}\,i\hgn(p\,')\,i\hgd(p_2\!-\!p\,')
\,u^c_2,
~~~~~ M^{(2)}_a=-M^{(1)}_a(N_1\LRA N_2) .
$$
Here: $u_i$ is the bispinor (isospinor also) of the $i$-th final nucleon, $\bar uu=2m$; 
$u^c=\tau_2 U_c\bar u^T=\tau_2\gamma_2 u^*$, where $U_c=\gamma_2\gamma_0$ is the charge-conjugation 
matrix; $\mgn^{(1,2)}=\bar u_{1,2}\hmgn^{(1,2)} u$ is the amplitude of subprocess $\gn_{1,2}$, and 
$u$ is the bispinor (isospinor also) of the intermediate nucleon with 4-momentum 
$\,p\,'\!=\!p_d-\!p_{\,2,1}$; ~$\hgn(p\,')=(p\sss^{\,\prime}\!+m)/(p\,'^{\,2}\!-m^2\!+i0)$ is the 
nucleon propagator, where $p\sss\equiv p_{\mu}\gamma_{\mu}$; $~\hgd(p_{\,2,1}-\!p\,')$ is the 
$dNN$-vertex related to the deuteron wave function (DWF) as given in Appendix~\ref{app:deut}.  The 
amplitude $M_a$ is antisymmetric with respect to the nucleon permutations in accordance with the 
Pauli principle.

Further, we retain only the positive-energy part of the nucleon propagator $G_N(p\,')$ and apply 
the connection between $\hgd$ and DWF $\hPsi$. Then, for a given spin and isospin states of the 
particles, we obtain
\be
	M^{(1)}_a=2\sqrt{m}\,\sum_{m',\tau'}
	\langle \pi, m_1,\tau_1 |\,\hmgn^{(1)}|\,\lambda, m',\tau'\rangle
	\langle m',\tau',m_2,\tau_2 |\,\hPsi(\bfp_2)|\,m_d\rangle,
\label{ma2}\ee
and the 2nd term is $M^{(2)}_a=-M^{(1)}_a$ (with permutation of the variables of the final nucleons). 
Here: $m_{1,2}$, $m'$, $\lambda$, and $m_d$ are spin states of the final nucleons, virtual nucleon, 
photon, and deuteron, respectively; $\pi$, $\tau_{1,2}$, and $\tau'$ are isospin states of pion, final 
nucleons, and virtual nucleon, respectively. Substituting isospin states for the reaction $\gdpp$, one 
gets
\be
	M^{}_a=2\sqrt{m}~\sum_{m'}~
	\bigl[\,\langle m_1|\,\hmgne^{(1)}|\,\lambda, m'\rangle
	\langle m',m_2 |\,\hPsi(\bfp_2)|\,m_d\rangle
\label{ma3}\ee
$$~~~~~~~~~~~~~~~~~~~~~~~
-\langle m_2|\,\hmgne^{(2)}|\,\lambda, m'\rangle
 \langle m',m_1|\,\hPsi(\bfp_1)|\,m_d\rangle\,\bigr],
$$
where now $\mgne^{(i)}=\langle m_i|\,\hmgne^{(i)}|\,\lambda,m'\rangle$ are the $\gnp_{\,i}$ amplitudes. 
The expressions for DWF  $\langle m_1,m_2\,|\hPsi(\bfp)|\,m_d\rangle$ are given in Appendix~\ref{app:deut}. 
The $\gn$ amplitudes $\hmgn$ can be expressed through the Chew-Goldberger-Low-Nambu (CGLN) 
amplitudes~\cite{CGLN} (see Appendix~\ref{app:gn}). The CGLN amplitudes as functions of the $\pi p_i$ 
invariant masses $W_i$ depend on the virtual nucleon momentum $p'$ through the relation 
$W^2_i=(q+p')^2=(k+p_i)^2$. Thus, the Fermi-motion is taken into account in the $\gdpp$ amplitude 
$M^{}_a$ (\ref{ma3}).  The matrix elements $\langle m_1|\,\hmgn|\,\lambda, m'\rangle$ are given in 
Appendix~\ref{app:agn}.

\subsection{NN final state interaction}
\label{sec:NN_FSI}

The NN-FSI term $M_b$ [Fig.~\ref{fig:g1}(b)] of the $\gd$ amplitude can be written in the form
\be
	M_b=-i\intdpp\sum_{m'_1,\,m'_2}\!\frac
	{\langle m'_1,m'_2 |\,\hmgd^{I\!A}|\,\lambda,m_d\rangle
	\langle m_1,m_2 |\,\hmm_{NN}|\,m'_1,m'_2\rangle}
	{\rtt (p\,'^{\,2}_1-m^2+i0)(p\,'^{\,2}_2-m^2+i0)}.
\label{nn1}\ee
Here: $m'_1$ and $m'_2$ are spin states of the intermediate nucleons; the notations for $m_1$, 
$m_2$, $\lambda$, and $m_d$ are the same as in Eqs.~(\ref{ma2}) (for short, we omit isospin 
indices); $\hmgd^{I\!A}$ is the amplitude of subprocess $\gd$ in impulse approximation
\be
	\langle m'_1,m'_2|\,\hmgd^{I\!A}|\,\lambda,m_d\rangle
	=2\sqrt{m}~\sum_{m'}\,\langle m'_1|\,\hmgn |\,\lambda, m'\rangle
	\langle m',m'_2 |\,\hPsi(\bfp_2^{\,\prime})|\,m_d\rangle,
\label{nn2}\ee
where $\hmm_{NN}$ is the $NN$-scattering amplitude.
The integral over the energy in Eq.~(\ref{nn1}) can be related to
the residue at the nucleon (momentum $p^{\,\prime}_{\,2}$) pole
with positive energy. Let us rewrite the 3-dimensional integral
$\int\!d\bfp^{\,\prime}_2$ in the $N\!N$ center-of-mass system.
Then, we get ~$p^{\,\prime\,2}_1\!-\!m^2\!+i0=\!2W(E\!-E'\!+i0)$,
where $W$ is the $N\!N$-system effective mass,
$E=\!W/2\!=\!\sqrt{p^{\,2}_N\!+\!m^2}$, $\pN\!=\!|\bpN|$,
$\,E'\!=\!\sqrt{p^{\,\prime\,2}_N\!+\!m^2}$, $\pNi\!=\!|\bpNi|$,
and $\,\bpN$($\bpNi$) is the relative 3-momentum in the final
(intermediate) $N\!N$ state. We obtain
\be
	M_b=\intdpi\,\frac{\langle\cdots\rangle}{\rtt 4E'\,W\,(E'-E-i0)},
\label{nn3}\ee
$$
\langle\cdots\rangle=\sum_{m'_1,\,m'_2}
\langle m'_1,m'_2 |\,\hmgd^{I\!A}|\,\lambda,m_d\rangle
\langle m_1,m_2 |\,\hmm_{NN}|\,m'_1,m'_2\rangle.
$$
One can rewrite $M_b$ as
\be
	M_b=M^{on}_b+M^{off}_b
	=\intdpi\,\frac{\langle\cdots\rangle}{4E'\,W}\,
	\left[i\pi\delta(E'\!-\!E)+P\,\frac{1}{E'\!-\!E}\right].
\label{nn4}\ee
Here, $M^{on}_b$ and $M^{off}_b$ are the contributions from the
1st and 2nd terms, respectively, in square brackets in the r.h.s.
of Eq.~(\ref{nn4}), where $P$ means the principal part of the
integral.  The amplitudes $M^{on}_b$ and $M^{off}_b$ correspond
to the on-shell and off-shell intermediate nucleons, respectively.
For $M^{on}_b$, we get
\be
	M^{on}_b=i\pi\!\intdpi\,\frac{\langle\cdots\rangle}{4E'\,W}
	\,\,\delta(E'\!-\!E)=\frac{i\pN}{32\pi^2 W}
	\,\int\!d\Omega'\,\langle\cdots\rangle,
\label{nn5}\ee
where $d\Omega'=dz'd\varphi'$ ($z'=\cos\theta'$) is the element
of solid angle of relative motion of the intermediate nucleons.
Consider the 2nd term $M^{off}_b$. Let us use Eqs.~(\ref{de1}) of
Appendix~\ref{app:deut} for $\hPsi(\bfp_2^{\,\prime})$ in
Eq.~(\ref{nn2}) and represent the integrand $\langle\cdots\rangle$
in Eq.~(\ref{nn3}) as the sum of two terms, proportional to S- and
D-wave components of DWF, i.e., $u(p'_2)$ and $w(p'_2)$. Then, we
obtain
\be
	\ba{c}
	\langle\cdots\rangle=A\,u(p'_2)+B\,w(p'_2)~~~ (p'_2=|\bfp'_2|),\\
	\rppt\dist
	A=2\sqrt{m}\!\sum_{m',m'_1,m'_2}\,
	\langle m'_1|\,\hmgn |\,\lambda,m'\rangle
	\langle m',m'_2 |\,\hat S_u|\,m_d\rangle,
	\langle m_1,m_2 |\,\hmm_{NN}|\,m'_1,m'_2\rangle,
\ea\label{nn6}\ee
and $B$ is given by the expression for $A$ after the replacement
$\hat S_u\!\to\!\hat S_w$, where $\hat S_u$ and $\hat S_w$ are
given in Eqs.~(\ref{de1}) of Appendix~\ref{app:deut}.  The
factors $A$ and $B$ contain $\gamma N$ and $N\!N$ amplitudes,
spin structure of DWF, and depend on the momenta of the particles
in Fig.~\ref{fig:g1}(b). Note that the $N\!N$-FSI amplitude
$M^{}_b$ (\ref{nn3}) takes into account the Fermi-motion, since
the amplitudes $\hmgn$ and $\hmm_{NN}$ depend on the intermediate
momenta $p'$ and $p'_2$, respectively.  In the integral
$\int\!d\bpNi=\int\!d\Omega'd\pNi p^{\,\prime\,2}_N$~(\ref{nn4}),
we take out of subintegral $\int\!d\pNi$ the factors $A$ and
$B$~(\ref{nn6}) at $\pNi=\pN$, i.e., we calculate $A$ and $B$ as
well as the amplitudes $\hmgn$ and $\hmm_{NN}$ with the on-shell
intermediate nucleons. This approximation means that we neglect
the off-shell dependence of the $\gamma N$ and $N\!N$ amplitudes
in comparison with sharp momentum dependence of DWF. Then, we
get
\be
	M^{off}_b=\ointpi\,\frac{\langle\cdots\rangle}{4E'\,W\,(E'\!-\!E)}\,
	=\frac{1}{32\pi^2 W}\int\!d\Omega'\,\bigl(A\,I_u+B\,I_w\bigr),
\label{nn7}\ee
$$
I_u=\ointp\,\frac{u(p'_2)}{E'\!-\!E},~~~~~~
I_w=\ointp\,\frac{w(p'_2)}{E'\!-\!E},
$$
where $\oint$ denotes the principal part of the integral. We also
include the formfactor $f(\pNi)$~\cite{Lev06} to parametrize the
off-shell ${^1}S_0$ partial amplitude of $pp$-scattering and define
the integrals
\be
	I^{(0)}_u=\ointp\frac{u(p'_2)f(\pNi)}{E'\!-\!E},~~~~~
	f(\pNi)=\frac{p^{\,2}_N+\beta^2}{p^{\,\prime\,2}_N+\beta^2}
\label{nn8}\ee
with $\beta=1.2\,$fm$^{-1}$~\cite{Lev06};
$I^{(0)}_w=I^{(0)}_u(u(p'_2)\!\to\! w(p'_2))$.  Let us write
the terms $A$ and $B$~(\ref{nn6}) as
\be
	A=A_0+A_1,~~~ B=B_0+B_1,
\label{nn9}\ee
where $A_0$ ($A_1$) is given by Eq.~(\ref{nn6}) when only
${^1}S_0$ part is saved (excluded) in the $pp$-scattering
amplitude $\hmm_{NN}$ (for $B_{0,1}$ the substitution $\hat
S_u\!\to\!\hat S_w$ in Eq.~(\ref{nn6}) is implied). Combining
Eqs.~(\ref{nn4})-(\ref{nn8}), we obtain
\be
	M_b=\int\!\frac{d\Omega'}{32\pi^2 W}\,
	\left[i\pN\bigl(A\,u(p'_2)+B\,w(p'_2)\bigr)+A^{}_0 I^{(0)}_u\!
	+\!A^{}_1 I^{}_u\!+\!B^{}_0 I^{(0)}_w\!+\!B^{}_1 I^{}_w\right].
\label{nn10}
\ee
The integrals $I^{}_u$, $I^{}_w$, $I^{(0)}_u$, $I^{(0)}_w$,
and $\,\int\!d\Omega'$~(\ref{nn10}) are carried out
numerically. The $N\!N$-scattering amplitude is described in
Appendix~\ref{app:NN}.

\subsection{$\pi N$ final state interaction}
\label{sec:piN_FSI}

The $\pi N$-FSI term $M_c$ [Fig.~\ref{fig:g1}(c)] of the $\gd$
amplitude can be written in the form
\be
	M_c=M^{(1)}_c+M^{(2)}_c,~~~~
	M^{(1)}_c=-\intdki\,\frac{\langle\cdots\rangle}
	{\rtt 2E'\,(k^{\,\prime\,2}\!-\mu^2\!+i0)},
\label{pin1}\ee
$$
\langle\cdots\rangle=\sum_{\pi',\tau'_2,m'_2}\langle
\pi',\tau_1,m_1,\tau'_2,m'_2|\,\hmgd^{I\!A}|\,\lambda,m_d\rangle\,
\langle \pi,\tau_2,m_2|\,\hmm^{(2)}_{\pi N}|\,\pi',\tau'_2,m'_2\rangle,
$$
where the integral over the energy is also related to the
residue at the nucleon pole (momentum $p^{\,\prime}_2$) as
in Eq.~(\ref{nn2}).  Here: $m'_2$ and $\tau'_2$ are spin
and isospin states of the intermediate nucleon with
4-momentum $p^{\,\prime}_2$; ~$\tau'$ is isospin states of
intermediate pion; the notations $m^{}_{1,2}$, $\tau^{}_{1,2}$,
$\pi$, $\lambda$, and $m_d$ are given above (see
Eq.~(\ref{ma2})); ~$M^{(2)}_{\pi N}=\langle\pi,\tau_2,m_2|\,
\hmm^{(2)}_{\pi N}|\,\pi',\tau'_2,m'_2\rangle$ is the
$\pi N\!\to\!\pi N_2$ amplitude; $\bki$ is the relative
3-momentum in the intermediate $\pi N$ system. The 2nd term
$M^{(2)}_c=-M^{(1)}_c$ (with permutation of the final
nucleons). Substituting isospin states for the reaction
$\gdpp$, and making use of Eq.~(\ref{nn2}), we get the
integrand $\langle\cdots\rangle$ in Eq.~(\ref{pin1}) in the
form
\be
	\langle\cdots\rangle=2\sqrt{m}\sum_{m',\,m'_2}
	\bigl[\,
	\langle m_1|\,\hmm^{(1)}(\gnp)|\,\lambda,m'\rangle
	\,\langle m_2|\,\hmm^{(2)}_{\pi^- p}|\,m'_2\rangle~~~~~~~~~
\label{pin2}\ee
$$~~~~~~
-\langle m_1|\,\hmm^{(1)}(\gpp)|\,\lambda,m'\rangle
\,\langle m_2|\,\hmm^{(2)}_{cex}|\,m'_2\rangle\bigr]\,
\langle m'\!,m'_2|\,\hPsi(\bfp'_2)|\,m_d\rangle,
$$
where $\hmm^{(i)}_{\pi N}$ and $\hmm^{(i)}_{cex}$ are the
elastic and charge-exchange ($\pi^0 n\!\to\!\pi^- p\,$ here)
$\pi N_i$ amplitudes, respectively. The relative sign ``-"
between two terms in Eq.~(\ref{pin2}) arises from isospin
antisymmetry of the DWF with respect to the nucleons. Further,
we rewrite the denominator $k^{\,\prime\,2}\!-\!\mu^2\!+\!i0$
in Eq.~(\ref{pin1}) as
\be
	k^{\,\prime\,2}\!-\!\mu^2\!+\!i0=2W_2(E-\!E'\!+\!i0),~~~
	E=\!\sqrt{k^{\,2}_2\!+m^2},~~~ E'\!=\!\sqrt{k'^{\,2}_2\!+m^2}
\label{pin3}\ee
($k_2=|\bfk_2|,~k'_2=|\bfk'_2|$),
where $W_2$ is the effective mass of the rescattering
$\pi N_2$ system, and $E$ ($E'$) is the total energy of
the final (intermediate) nucleon in the $\pi N_2$ rest
frame.  In a way similar to Subsection~\ref{sec:NN_FSI},
we split the amplitude $M^{(1)}_c$ into ``on-shell" and
``off-shell" parts, and obtain
\be
	M^{(1)}_c\!=M^{(1),\,on}_c\!+M^{(1),\,off}_c\!
	=\int\!\!\frac{d\Omega'}{32\pi^2 W_2}\,
	\left[A\,\bigl(ik_2\,u(p'_2)+I_u\bigr)
	+B\,\bigl(ik_2\,w(p'_2)+I_w\bigr)\right],
\label{pin4}\ee
$$
I_u=\ointk\,\frac{u(p'_2)}{E'\!-\!E},~~~~~~~
I_w=\ointk\,\frac{w(p'_2)}{E'\!-\!E}.
$$
Here: $d\Omega'=dz'd\varphi'$ is the element of solid angle
of relative motion in the intermediate $\pi N$ system; the
factor $A$($B$) is given by the r.h.s. of Eq.~(\ref{pin2})
after the replacement $\hPsi(\bfp'_2)\!\to\!\hat S_u(\hat S_w)$
(see Appendix~\ref{app:deut}, Eq.~(\ref{de1})), and is calculated
with the on-shell intermediate pion and nucleon.  The ``off-shell"
part $M^{(1),\,off}_c$ of the amplitude $M^{(1)}_c$~(\ref{pin4})
is given by the terms, containing the integrals $I_u$ and $I_w$.
The $\pi N$-scattering amplitude is described in
Appendix~\ref{app:piN}.

\section{Results}
\label{sec:results}

\subsection{Comparison with the experiment}
\label{sec:compar}

\begin{figure}
\begin{center}
\begin{minipage}[b]{15cm}
\includegraphics[width=14cm, keepaspectratio]{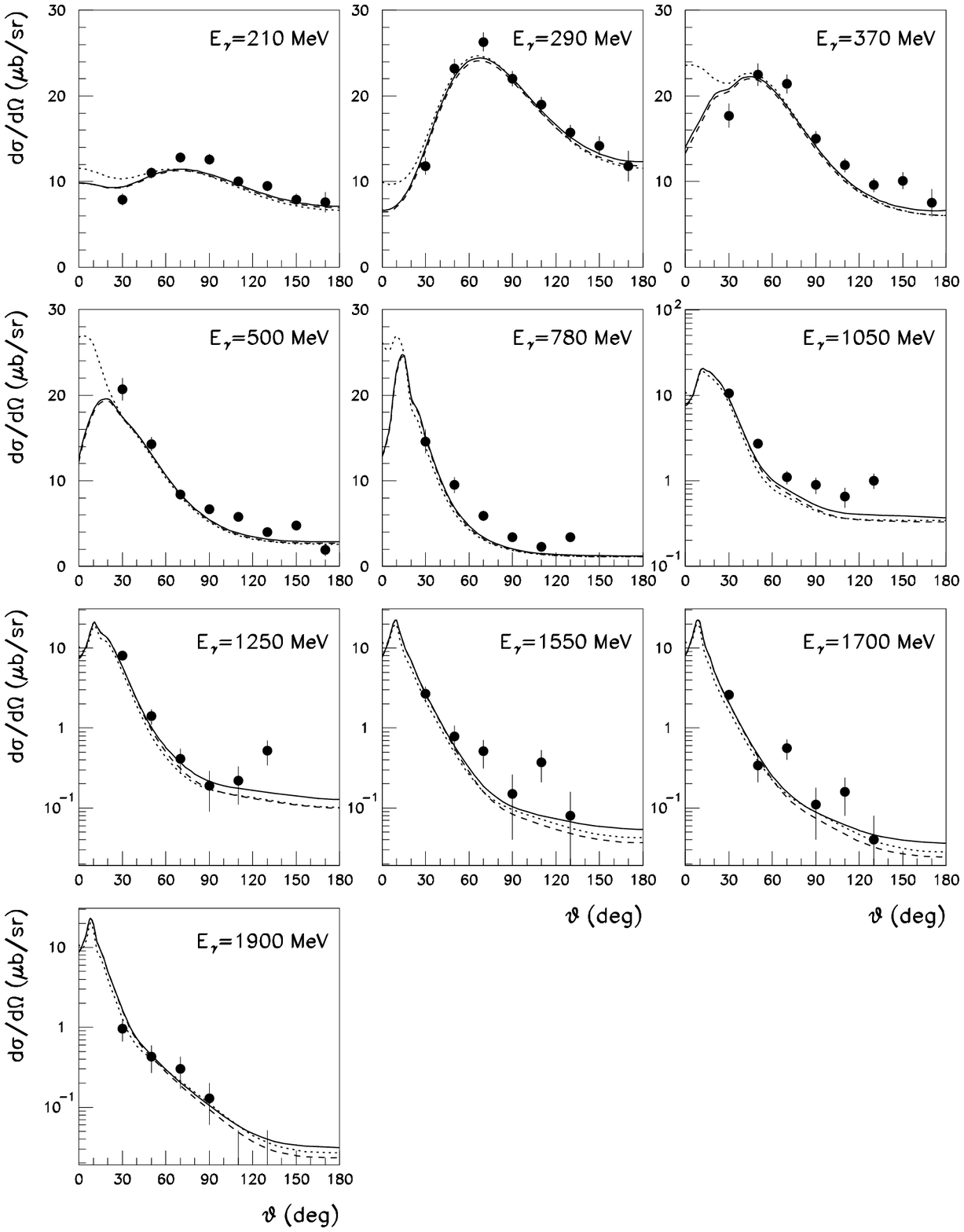}
\end{minipage}
\end{center}
\caption{The \dif\ \crs\ $\dso$ of the reaction $\gdpp$ in the
    laboratory frame at different values of the photon laboratory 
    energy $\ega\le$1900~MeV; $\theta$ is the polar angle of
    the outgoing $\pi^-$. Dotted curves show the contributions
    from the IA amplitude $M_a$ [Fig.~\protect\ref{fig:g1}(a)].
    Successive addition of the NN-FSI
    [Fig.~\protect\ref{fig:g1}(b)] and $\pi N$-FSI
    [Fig.~\protect\ref{fig:g1}(c)] amplitudes leads to
    dashed and solid curves, respectively. The filled
    circles are the data from Ref.~\protect\cite{Benz}.}
    \label{fig:g2}
\end{figure}

We present herein the results of calculations and comparison with
the experimental data on the \dif\ \crss\ $d\sgd(\theta)/d\Omega$,
where $\Omega$ and $\theta$ are solid and polar angles of outgoing
$\pi^-$'s in the laboratory frame, respectively, with z-axis along
the photon beam. The results are given in Fig.~\ref{fig:g2} for a
number of the photon energies $\ega$. Calculations were done with
DWF of the CD-Bonn potential (full model)~\cite{BonnCD}. The
filled circles denote the data from the bubble chamber experiment
at DESY~\cite{Benz}.

\begin{figure}
\begin{center}
\begin{minipage}[b]{15cm}
\includegraphics[width=14cm, keepaspectratio]{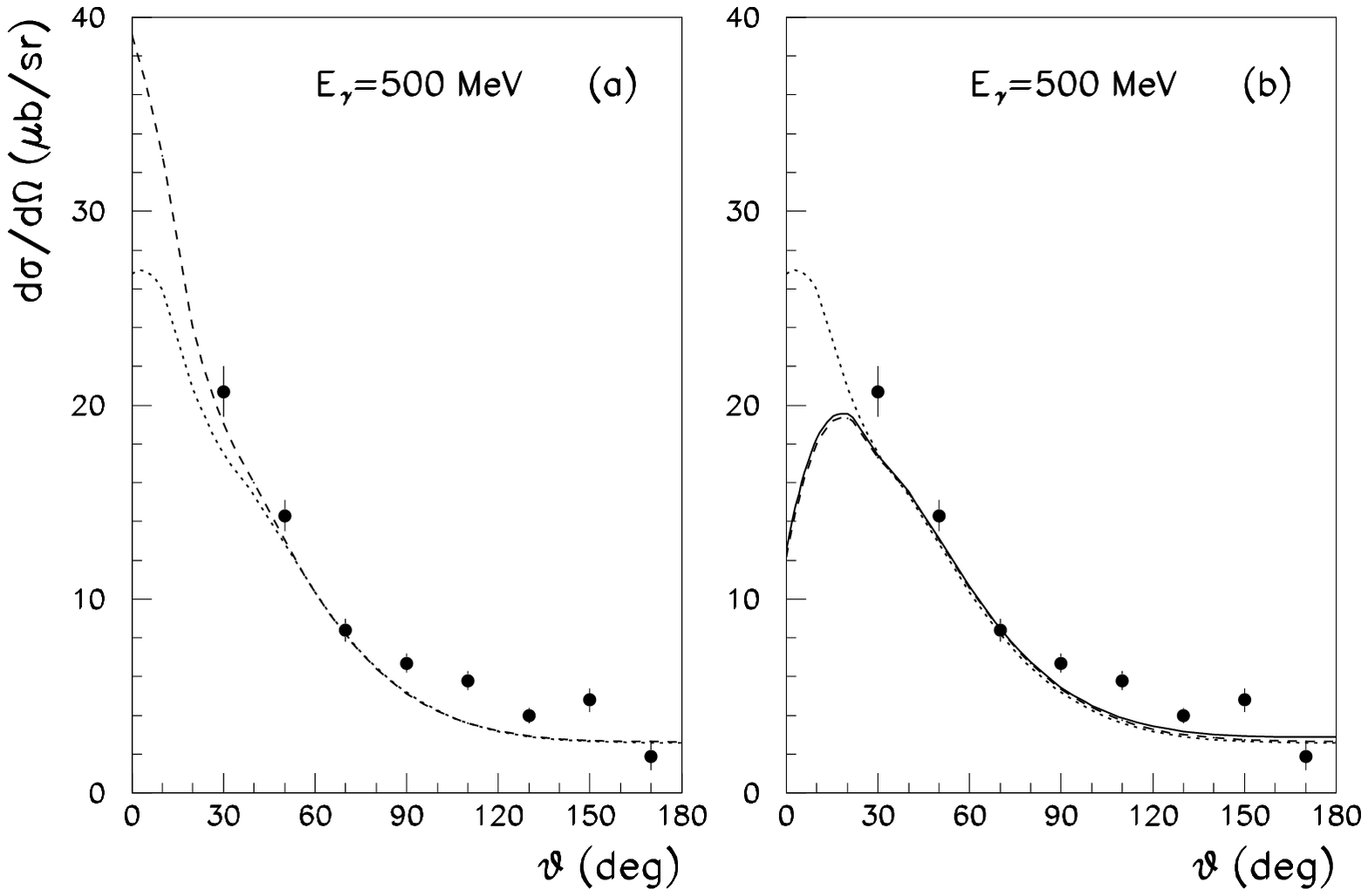}
\end{minipage}
\end{center}
\caption{The \dif\ \crs\ of the reaction $\gdpp$ in the laboratory frame
    at $\ega=500$~MeV. (a): the dotted curve is the IA
    contribution, i.e., the same as in Fig.~\protect\ref{fig:g2};
    the dashed one is the contribution from
    $|M^{(1)}_a|^2\!+\!|M^{(1)}_b|^2$, i.e., without the
    cross term.  (b): the solid and dashed curves means the
    same as in Fig.~\protect\ref{fig:g2}; the dotted curve
    is obtained with the IA-term and $S$-wave part of NN-FSI.
    The data are from Ref.~\protect\cite{Benz}.}
    \label{fig:g4}
\end{figure}

The dotted curves show the results obtained with the IA
amplitude $M_a$ [Fig.~\ref{fig:g1}(a)]. It is known that
the IA \crs\ $\sigma(\gdpp)$ can be expressed in the closure
approximation~\cite{Chew} through the \crs\ $\sigma(\gnp)$
and Pauli correction factor, which comes from the cross term
of the amplitudes $M^{(1)}_a$ and $M^{(1)}_b$~(\ref{ma1}).
It reads
\be
    \dsdo(\gdpp)=\dsdo(\gnp)\Bigr[1-F^{}_S(\Delta)+[\cdots]
    \frac{\overline{|\bfkk|^2}}{\overline{|L|^2}+\overline{|\bfkk|^2}}
    \,F^{}_S(\Delta)\Bigl] .
\label{c1}\ee Here: $[\cdots]$ is the Pauli factor,
$F^{}_S(\Delta)$ is the spherical form factor of the deuteron (we
neglect the contribution of the quadrupole form factor), and
$\Delta=\bfp_1\!+\bfp_2$ is 3-momentum transfer;
$\overline{|L|^2}$ and $\overline{|\bfkk|^2}$ are non-spin flip
and spin flip $\gnp$ amplitudes, respectively (see
Appendix~\ref{app:agn}, Eq.~(\ref{agn1})) squared and averaged
over the photon polarization. For zero-angle ($\theta\!=0$) pions,
the non-spin flip term $\overline{|L|^2}=0$. Then at
$\Delta\!\to\!0$, we have $F^{}_S(\Delta)\!\to\!1$ and the Pauli
factor $[\cdots]\!\to\!2/3$ in Eq.~(\ref{c1}). At
$\Delta\to\infty$, we have $F^{}_S(\Delta)\!\to\!0$ and
$[\cdots]\!\to\!1$. The momentum transfer $\Delta$ increases
together with the laboratory angle $\theta$.  Thus, the spectra on
Fig.~\ref{fig:g2} should be partly suppressed at small angles
$\theta\sim 0$ as compared with $\dso(\gnp)$. Fig.~\ref{fig:g4}(a)
shows two different results for $d\sgd(\theta)/d\Omega$ at
$\ega=500$~MeV: the dotted curve represents the contribution from
the IA amplitude squared $|M^{(1)}_a\!+\!M^{(1)}_b|^2$ and the
dashed one shows the contribution from
$|M^{(1)}_a|^2\!+\!|M^{(1)}_b|^2$, i.e., without the cross term.
The difference of the curves in Fig.~\ref{fig:g4}(a), i.e., the
Pauli effect, at small angles is clearly seen.

The dashed curves in Fig.~\ref{fig:g2} are
the contribution of IA- and NN-FSI terms $M_a\!+M_b$
[Fig.~\ref{fig:g1}(a),(b)]. The solid curves show the results
obtained with the full amplitude $M_a\!+M_b\!+M_c$
[Fig.~\ref{fig:g1}(a),(b),(c)], including IA-, NN-, and
$\pi$N-FSI terms. Fig.~\ref{fig:g2} shows a sizeable FSI effect
at small angles $\theta\lesssim 30^{\,\circ}$, and it mainly
comes from NN-FSI (the difference between dotted and dashed
curves). Comparing dashed and solid curves, one finds that
$\pi$N-FSI affects the results very slightly. Note that at the
energies $\ega=300-500$~MeV, the effective masses of the final
$\pi p$ states predominantly lie in the $\Delta(1232)3/2^+$ region.
Thus, the plots at $\ega=370$ and 500~MeV of Fig.~\ref{fig:g2}
show that the role of $\pi$N-rescattering even in the
$\Delta(1232)3/2^+$ region is very small.

\begin{figure}
\begin{center}
\begin{minipage}[b]{15cm}
\includegraphics[width=14cm, keepaspectratio]{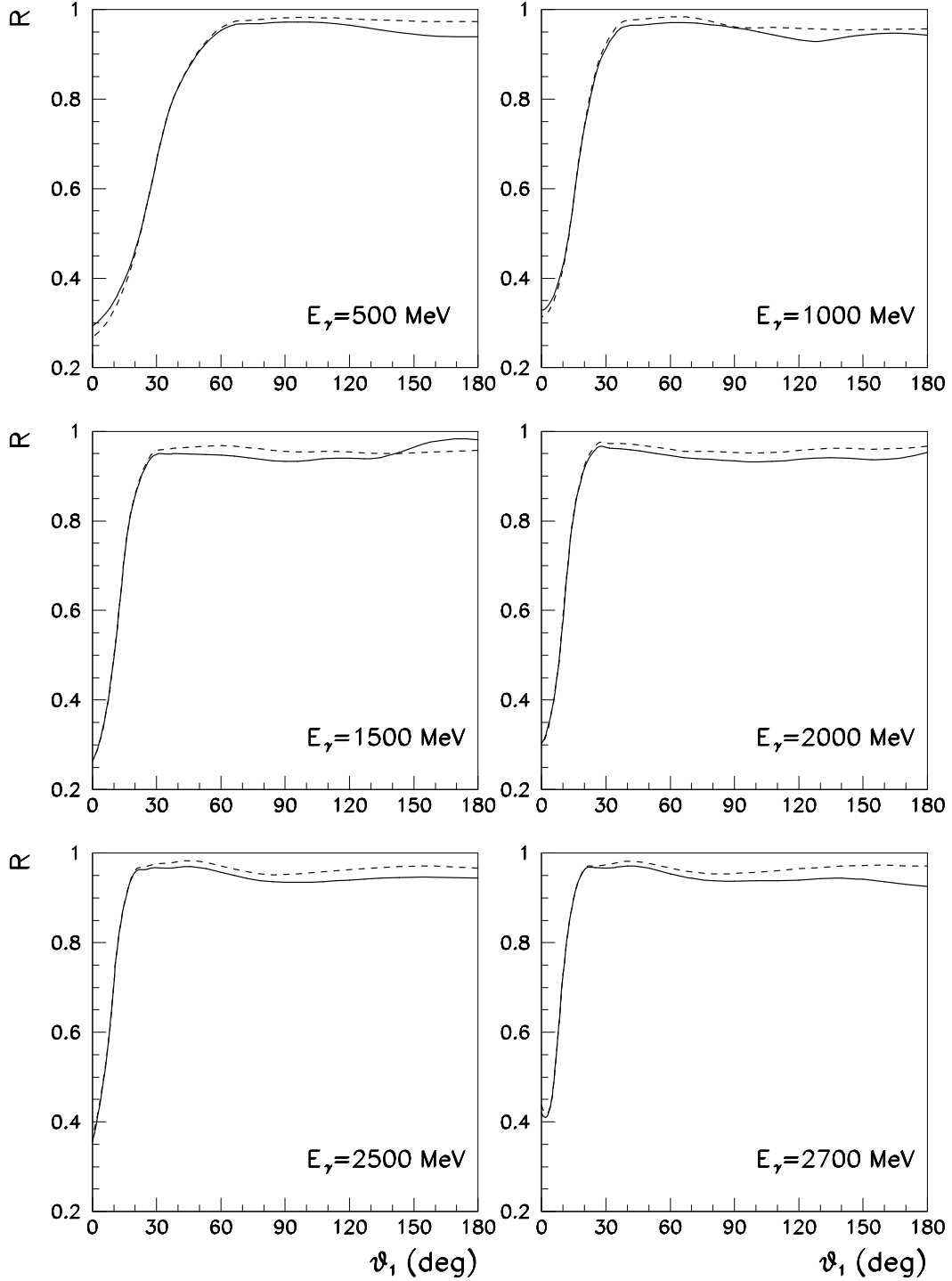}
\end{minipage}
\end{center}
\caption{The correction factor $R$, defined by Eq.~(\protect\ref{ne7}),
    where $\theta_1$ is the polar angle of the outgoing $\pi^-$ in
    the rest frame of the pair $\pi^-+$fast proton. The kinematical
    cut~~(\ref{cut}) is applied.  The solid (dashed) curves are
    obtained with both $\pi N$- and NN-FSI (only NN-FSI), taken
    into account.} \label{fig:g5}
\end{figure}

Fig.~\ref{fig:g5} demonstrates a reasonable description of
the data~\cite{Benz} on $d\sgd(\theta)/d\Omega$. These data
are also confirmed by recent results from the GDH
experiments~\cite{Ahrens} in Mainz. Note that the data are
absent at small angles $\theta\lesssim 30^\circ$, where the
FSI effects are sizeable. This is also the region of the
most pronounced disagreements between the theoretical
predictions of different authors~\cite{Ahrens}.

The role of FSI is shown in more detail at $\ega=500$~MeV in
Fig.~\ref{fig:g4}(b). Here, the dashed curve is the result obtained
including the IA-term and the $S$-wave part of NN-FSI. The dotted
and solid curves mean the same as in Fig.~\ref{fig:g2}. Thus we
see that at small angles, the $S$-wave part of NN-FSI dominates the
FSI contribution.

At large angles, the FSI effects are more significant as the photon
energy increases. It is evident from the plots at $\ega\ge
1050$~MeV in Fig.~\ref{fig:g2}. Our interpretation is that at both
high energies and at large angles, the role of configurations with
fast final protons increases. For these configurations, the IA
amplitude is suppressed by the deuteron wave function in
comparison to the rescattering terms. These kinematical regions
were considered in more detail in Ref.~\cite{La06}.

\subsection{Extraction of the $\gnp$ \crss\ from the $\gamma d$
    data}
\label{sec:extract}

The data on the deuteron target does not provide direct
information on the \dif\ \crs\ $\dso(\gnp)$, because of the
$\gdpp$  squared amplitude term $\overline{|\mgd|^2}$, where
$\mgd=M_a+M_b+M_c$ [Fig.~\ref{fig:g1}] and can not be expressed
directly through the term $\overline{|M_{\gamma n}|^2}$. Let us
neglect for the time the FSI amplitudes,
[Fig.~\ref{fig:g1}(b),(c)] and let the final proton with momentum
$\bfp_1$($\bfp_2$) be fast (slow) in the laboratory system and
denoted by $p_1$($p_2$). Then, the IA diagram $M^{(1)}_a$ with a
slow proton $p_1$ emerging from the deuteron vertex dominates,
$M^{(2)}_a$ is suppressed, and $\mgd\approx M^{(1)}_a$.  This
approximation corresponds to the ``quasi-free" (QF) process on the
neutron. In this case, one can relate the \dif\ \crs\
$\dso_1(\gnp)$ on neutron with that on the deuteron target as
follows.  (Hereafter, $\Omega_1$ is the solid angle of relative
motion in the $\pi^-p_1$ pair.)  From Eq.~(\ref{ma2}), we get \be
    \overline{|M^{(1)}_a|^2}=4m\,\overline{|\mgne^{(1)}|^2}
    (2\pi)^3\rho(p_2),~~~
    (2\pi)^3\rho(p)=u^2(p)+w^2(p),~~~ \int\!\rho(p)\,d\bfp=1,
\label{ne1}\ee where $\rho(p)$ is the momentum distribution in the
deuteron. Making use of Eqs.~(\ref{kin1}) and (\ref{ne1}), and
multiplying by a factor of 2 (we include also the configuration 
when slow and fast protons are replaced, and the amplitude $M^{(2)}_a$
dominates), we obtain
\be
    \frac{d\sgd^{QF}}{d\bfp_2\,d\Omega_1}=
    n(\bfp_2)\,\frac{d\sgn}{d\Omega_1},~~~
    n(\bfp_2)=\frac{\ega^{\,\prime}}{\ega}\,\rho(p_2),
    ~~~\frac{\ega^{\,\prime}}{\ega}=1+\beta\cos\theta_2,~~~
    \beta=\frac{p_2}{E_2}
\label{ne2}\ee (see, for example, Refs.~\cite{BL,La78,La81}).
Here: $\ega^{\,\prime}$ is the photon energy in the rest frame of
the virtual neutron with momentum $p'$ [Fig.~\ref{fig:g1}(a)]; the
factor $\ega^{\,\prime}/\ega$ is the ratio of photon fluxes in
$\gamma d$ and $\gamma n$ reactions; $\theta_2$ is the laboratory polar
angle of final slow proton $p_2$. Hereafter, we use the notation
$d\sgd^i/d\bfp_2d\Omega_1$, where index ``$i$" specifies the
$\gdpp$ amplitude $\mgd^{\,i}$, namely $\,\mgd^{QF}\!=M^{(1)}_a$
and $\mgd^{IA}\!=M_a$. The notation $d\sgd/d\bfp_2d\Omega_1$
(without index) represents the \dif\ \crs, calculated according to
Eqs.~(\ref{kin1}) with full amplitude $\mgd=M_a+M_b+M_c$.  Let us
rewrite Eqs.~(\ref{ne2}) in the form
\be
    \frac{d\sgd}{d\bfp_2 d\Omega_1}=n(\bfp_2)\,r\,\frac{d\sgn}{d\Omega_1},
    ~~~r=r^{}_P\,r^{}_{FSI},~~~ r^{}_P=\frac{({\rm IA})}{({\rm QF})},~~~
    r^{}_{FSI}=\frac{({\rm full})}{({\rm IA})},
\label{ne3}\ee
where for short, we use the notations (full)$\,=d\sgd/d\bfp_2d\Omega_1$
and $(i)=d\sgd^i/d\bfp_2d\Omega_1$ for $i=\,$QF and IA.  Eqs.~(\ref{ne3})
enable one to extract the \dif\ \crs\ $d\sgn/d\Omega_1$ on neutron from
$d\sgd/d\bfp_2d\Omega_1$, making use of the factors $n(\bfp_2)$ and $r$.
Here: the factor $n(\bfp_2)$, defined in Eqs.~(\ref{ne2}), takes into
account the distribution function $\rho(p_2)$ and Fermi-motion of
neutron in the deuteron;
$r=r^{}_P\,r^{}_{FSI}$ is the correction coefficient, written as the
product of two factors of different nature. The factor $r^{}_P$ takes
into account the difference of IA and QF approximations. Formally, we
call it ``Pauli correction" factor, since the IA amplitude
$M^{}_a=M^{(1)}_a\!+M^{(2)}_a$ is antisymmetric over the final nucleons.
However, the factors $r^{}_P$ in Eqs.~(\ref{ne3}) and the expression
in square brackets $[\cdots]$ in Eq.~(\ref{c1}) are not identical. The
factor $r^{}_{FSI}$ in Eqs.~(\ref{ne3}) is the correction for
``pure" FSI effect.

Generally for a given photon energy $\ega$, the \crs\
$d\sgd/d\bfp_2d\Omega_1$~(\ref{ne3}) with unpolarized particles
and the factor $r$ depend on $p_2$, $\theta_2$, $\theta_1$, and
$\varphi_1$ (4 variables), where $\theta_1$ and $\varphi_1$ are
the polar and azimuthal angles of relative motion in the final
$\pi^-p_1$ pair.  To simplify the analysis, we integrate the \dif\
\crs\ on deuteron over $\bfp_2$ in a small region $p_2<p_{max}$
and average over $\varphi_1$. Then, we define \be
    \frac{d\sgd^i}{d\Omega_1}(\ega,\theta_1) =\frac{1}{2\pi}
    \int\!\frac{d\sgd^i}{d\bfp_2\,d\Omega_1}\,\,d\bfp_2 d\varphi_1,
\label{ne4}\ee where the index ``$i$" was introduced above (after
Eqs.~(\ref{ne2})). The \crs~(\ref{ne4}) depends on $\ega$ and
$\theta_1$. We calculate the same integral from the r.h.s of
Eqs.~(\ref{ne2}). Then, we take the \crs\ $\sgn/d\Omega_1$ out of
the integral $\int\!d\bfp_2$, assuming $n(\bfp_2)$ to be a sharper
function.  Thus, making use of Eqs.~(\ref{ne2})-(\ref{ne4}), we
obtain \be
    \frac{d\sgd^{QF}}{d\Omega_1}(\ega,\theta_1)=
    c\,\frac{d\asgn}{d\Omega_1},~~~
    c=\int\!n(\bfp_2)\,d\bfp_2~~~ (|\bfp_2|<p_{max}),
\label{ne5}\ee
where $d\asgn/d\Omega_1$ is averaged over the energy $\ega'$ in some
region $\ega'\sim\ega$. The value $c=c(p_{max})$ can be called the
``effective number" of neutrons with momenta $p<p_{max}$ in the
deuteron.  Under the restriction $|\bfp_2|<p_{max}$ in the integral
for $c$~(\ref{ne5}), we get
\be
    c(p_{max})=4\pi\!\!\int\limits^{p_{max}}_0\!\!\rho(p)p^2 dp
    \to 1 ~~~{\rm at}~~p_{max}\!\to\!\infty.
\label{ne6}\ee
A number of values of $c(p)$ are given in the Table~\ref{tab:tbl1} for
two versions of CD-Bonn DWF~\cite{BonnCD}.
\begin{table*}[ht]
\caption{``Effective number" of neutrons with momenta $p<p_{max}$ in
    the deuteron. \label{tab:tbl1}}
\begin{center}
\begin{tabular}{|c|c|c|c|c|c|}
    \hline\rpt $\,p_{max}$ (MeV$\!/c$) & 50 & 100 & 200 & 300 & Ref. \\
    \hline\rpt $c(p_{max})$ & ~0.335~ & ~0.719~ & ~0.941~ & ~0.981~ &
    (full model)~\protect\cite{BonnCD}\\
    \hline\rpt $c(p_{max})$ & ~0.326~ & ~0.704~ & ~0.932~ & ~0.978~ &
    (energy-independent)~\protect\cite{BonnCD}\\
    \hline
\end{tabular}\end{center}
\end{table*}

Further, we rewrite Eqs.~(\ref{ne5}) in the form
\be
    \frac{d\sgd}{d\Omega_1}(\ega,\theta_1)=c\,R\,\frac{d\asgn}{d\Omega_1},
    ~~~ R=R^{}_P\,R^{}_{FSI},~~
    R^{}_P=\frac{({\rm IA})}{({\rm QF})},~~
    R^{}_{FSI}=\frac{({\rm full})}{({\rm IA})}.
\label{ne7}\ee
Here: $(i)=d\sgd^i/d\Omega_1$ ($i=\,$QF and IA) and (full)$=d\sgd/d\Omega_1$
(the definitions are different from those in Eqs.~(\ref{ne3})); the
factors $R$, $R^{}_P$, and $R^{}_{FSI}$ are similar to $r$, $r^{}_P$, and
$r^{}_{FSI}$, respectively, but defined as the ratios of the ``averaged"
\crss\ $d\sgd^i/d\Omega_1$.

Finally, we replace $d\sgd/d\Omega_1$ in Eqs.~(\ref{ne7}) by the $\gdpp$
data and obtain
\be
    \frac{d\asgn^{\,exp}}{d\Omega_1}(\bar\ega,\theta_1)=c^{-1}\!(p_{max})
    \,R^{-1}\!(\ega,\theta_1)\,
    \frac{d\sgd^{exp}}{d\Omega_1}(\ega,\theta_1),
\label{ne8}\ee where $d\asgn^{\,exp}/d\Omega_1$ is the neutron
\crs, extracted from the deuteron data $d\sgd^{\,exp}/d\Omega_1$.
Since the factor $R=$(full)/(QF) is the ratio of the calculated
\crss, we assume that (full)$\,\equiv
d\sgd^{theor}/d\Omega_1=d\sgd^{exp}/d\Omega_1$. The factor $R$ in
Eq.~(\ref{ne6}) is the function of the photon laboratory energy $\ega$
and pion angle $\theta_1$ in the $\pi^-p_1$ frame, but also
depends on the kinematical cuts applied.  The value $\bar\ega$ in
Eq.~(\ref{ne8}) is some ``effective" value of the energy
$\ega'=\ega(1+\beta\cos\theta_2)$ in the range $\ega(1\pm\beta)$.
Limiting the momentum $p_2$ to small values, we have $\beta\ll 1$
and $\bar\ega\approx\ega$. This approximation also improves, since
$\rho(p_2)$ peaks at $p_2=0$, where $\ega'=\ega$.

Eq.~(\ref{ne8}) is implied to be self-consistent, i.e., the $\gnp$
amplitude, extracted from the $d\asgn^{exp}/d\Omega_1$ is the same
as that used in calculations of the correction factor $R$. Then,
the following iterations are proposed. The 1st step: one obtains
the \crs\ $d\asgn^{exp}/d\Omega_1$ from Eq.~(\ref{ne8}) at $R=1$
(no corrections), making use of the coefficient $c(p_{max})$, and
extracts the $\gamma n$ amplitude $\mgn^{(0)}$ (0th approximation).
The next step: one calculates the factor $R$ defined in
Eqs.~(\ref{ne7}), making use of the amplitude $\mgn^{(0)}$ for the
calculations of the \crss\ $d\sgd^i/d\Omega_1$. Then, we repeat the
procedure of the previous step with new value of $R$, and obtain
the amplitude $\mgn^{(1)}$ in the 1st approximation. The procedure
can be continued.  If the correction is small, i.e., $R\approx 1$
($|R-1|\ll 1$), then $\mgn^{(1)}$ is a good approximation for the
corrected $\gnp$ amplitude.  Since, there are regions, where
$R\sim 1$ the FSI effects are insignificant and the preliminary
analysis of the $R$ factor is important for the procedure of the
extraction of the $\gnp$ amplitudes.

\subsection{Numerical results for the $R$ factor}
\label{sec:r-factor}

We present the results, obtained with the model discussed above,
for the correction factor $R$, defined in Eqs.~(\ref{ne7}). The
results depend on the kinematical cuts. We use cuts, similar to
those applied to the CLAS data events~\cite{Wei}, and select
configurations with
\be
    |\bfp_2|<200~MeV/c<|\bfp_1|,
\label{cut}\ee where $\bfp_1(\bfp_2)$ is the 3-momentum of fast
(slow) final proton in the laboratory system. The results are
given in Fig.~\ref{fig:g5} as functions of the photon laboratory
energy $\ega$ and $\theta_1$, where $\theta_1$ is the polar angle
of outgoing $\pi^-$ in the $\pi^- p_1$ rest frame with z-axis
directed along the photon momentum.

The solid curves show the results for $R$, where the \dif\ \crs\
(full) in Eqs.~(\ref{ne5}) takes into account the full amplitude
$M_a\!+M_b\!+M_c$ [Fig.~\ref{fig:g1}(a),(b),(c)]. The dashed curves
were calculated, excluding the $\pi$N-FSI contribution from the
(full) \crs. The main features of the results in Fig.~\ref{fig:g5}
are
\begin{enumerate}
\item a sizeable effect is observed in the region close to
    $\theta_1\!=0$, which narrows as the energy $\ega$ increases;
\item the correction factor $R$ is close to 1 (small effect) in the
    larger angular region.
\end{enumerate}

\begin{figure}
\begin{center}
\begin{minipage}[b]{15cm}
\includegraphics[width=12cm, keepaspectratio]{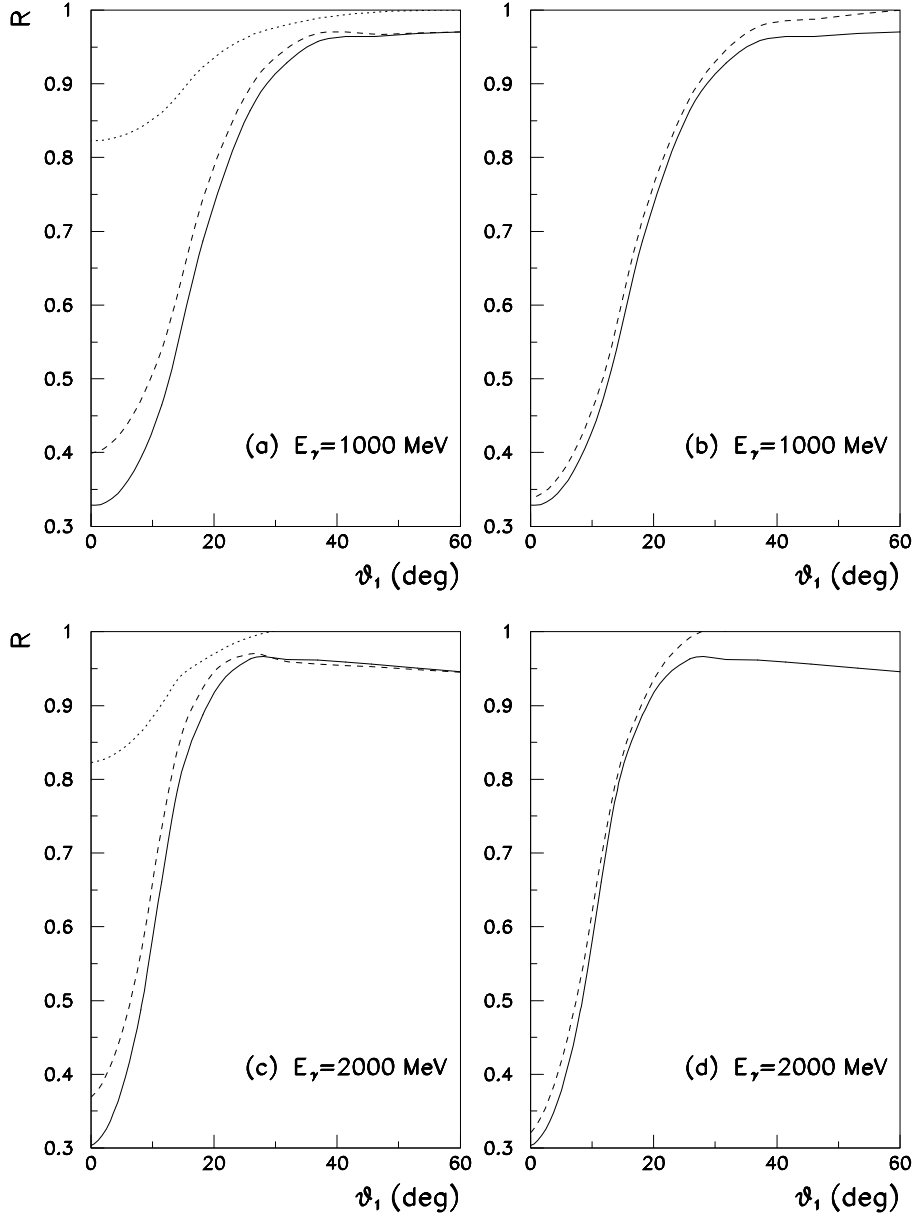}
\end{minipage}
\end{center}
\caption{The correction factors at $\ega=1000$~MeV [(a),(b)] and
        $\ega=2000$~MeV [(c),(d)]. The solid curves are the same
        as in Fig.~\protect\ref{fig:g5}.
        (a) and (c): the dashed (dotted) curves are the results
    for the factor $R^{}_{FSI}$($R^{}_P$), defined in
    Eq.~(\protect\ref{ne7}).
        (b) and (d): the dashed curves show the factor $R$, when
        $R^{}_{FSI}$ takes into account only the $S$-wave part of
    NN-FSI.} \label{fig:g6}
\end{figure}

Since $R$ consist of two factors $R^{}_P$ and $R^{}_{FSI}$, we also
present them separately in Figs.~\ref{fig:g6}(a) and ~\ref{fig:g6}(c)
for $\ega=1000$ and 2000~MeV, respectively. Here: dotted, dashed, and
solid curves show the values of $R^{}_P$, $R^{}_{FSI}$, and $R$,
respectively; the factor $R_{FSI}$ was calculated with the full
amplitude $M_a\!+M_b\!+M_c$ [Fig.~\ref{fig:g1}(a),(b),(c)] taken
into account. We find that $R^{}_P\ne 1$ at small angles, i.e., the
factor $R^{}_P$ in addition to the pure FSI factor $R^{}_{FSI}$
also contributes to the total correction factor $R$.

This can be naturally understood. Since $R^{}_P$ is the correction
for the 2nd (``suppressed") IA amplitude $M^{(2)}_a$, one should
expect $M^{(1)}_a\sim M^{(2)}_a$ and $R^{}_P\ne 1$ at
$\bfp_1\sim\bfp_2$.  The probability of such configuration
increases at $\theta_1\!\to\!0$.
It is clear that the possibility of the configuration
$\bfp_1\sim\bfp_2$ and the value of $R^{}_P$ should be rather
sensitive to kinematical cuts.

The dominant role of the $S$-wave NN rescattering in the FSI
effect was marked in Subsection~\ref{sec:compar}. This
contribution to the factor $R$ is presented in
Figs.~\ref{fig:g6}(b) and ~\ref{fig:g6}(d) for $\ega=1000$ and
2000~MeV, respectively. Here, solid curves mean the same as in
Fig.~\ref{fig:g5}, i.e. the total results; the dashed curves show
the values $R$, where $R^{}_{FSI}$ takes into account only the
correction from the $S$-wave part of NN-FSI. Comparing the solid
and dashed curves, we see that the FSI effect mostly comes from
the $S$-wave part of $pp$-FSI.  Note that the $S$-wave $pp$
amplitude and the total elastic $pp$ \crs\ $\sigma_{el}(pp)$
sharply peak near the threshold at the relative momentum
$p^{}_N\approx 23$~MeV$/c^2$. Thus, the $S$-wave NN-FSI effect
should be important in some region $\bfp_1\sim\bfp_2$, i.e., at
small angles as mentioned above and is evident from
Figs.~\ref{fig:g6}(b) and ~\ref{fig:g6}(d). Obviously, the result
is sensitive to the kinematical cuts.

\subsection{Factor $R$ and Glauber approximation}
\label{sec:glauber}

Now consider the region of large angles $\theta_1$, where FSI
effects are small ($R\sim 1$). In this case, we have the
rescattering of fast pion and nucleon on the slow
nucleon-spectator with small momentum transfer. Then, we may
estimate the FSI amplitudes in the Glauber approach~\cite{Frank},
if the laboratory momentum of the rescattered particle $\gg \bar
p$ (typical value in deuteron). For $N\!N$-FSI, this condition
gives $\sin\theta_1\gg{\bar p}W_1/m\ega$, where $W_1$ is the
$\pi^-p_1$ effective mass. Taking $\bar p=150$~MeV$/c$, we get
$\theta_1\!\gg 15.4^{\,\circ}(10^{\,\circ})$ for
$\ega\!=1000~(2000)$~MeV. As for the $\pi N$-FSI, we should also
exclude some region close to $\theta_1\sim 180^{\,\circ}$, where
$\pi^-$ is slow in the laboratory system. The high-energy
$NN$-scattering amplitude can be written as
\be
    M^{t}_{N\!N}=2ip\,W\sigma^t_{N\!N}\exp(bt),~~~
\label{r1}\ee where $p$, $W$, $t$, $b$, and $\sigma^t_{N\!N}$ are
the relative momentum, $N\!N$ effective mass, square of the
4-momentum transfer, slope, and total $N\!N$ \crs, respectively.
The amplitude is assumed to be purely imaginary, and spin-flip
term is neglected. Retaining only the $S$-wave part of DWF, we
obtain the IA- and NN-FSI amplitudes ($M_a$ and $M_b$) in the form
\be
    M_a=M^{(1)}_a=\langle\cdots\rangle\,u(p_2),~~~~~
    M_b=-\frac{1}{4}\,\sigma^t_{N\!N}\,\langle\cdots\rangle\, J,
\label{r2}\ee
$$
\dist
\langle\cdots\rangle=2\sqrt{m}\sum_{m'}\,\,
\langle m_1|\,\hmgn |\,\lambda,m'\rangle
\langle m',m'_2 |\,\hat S_u|\,m_d\rangle,~~~
J=\!\int\!\frac{d^2p_{\perp}}{(2\pi)^2}\,u(p_{\perp}\!)\,e^{bt}.
$$
Here, the IA amplitude $M^{(1)}_a$ is equal to the 1st term in the
r.h.s. of Eq.~(\ref{ma3}) with the replacement $\hPsi(\bfp_2)\to
u(p_2)\hat S_u$ (see Eqs.~(\ref{de1})); the 2nd term ($M^{(2)}_a$)
of the IA amplitude is neglected;
$t=-b(\bfp_{2\perp}\!-\!\bfp_{\perp})^2$, where
$\bfp_{2\perp}\!(\bfp_{\perp})$ is the transverse 2-momentum of
slow final (intermediate) proton with $Oz\parallel\bfp_1$
(fast-proton momentum). The factor $\exp(bt)$ is smooth in
comparison with sharper DWF $u(p_{\perp})$ in the integral
$J$~(\ref{r2}); thus, we neglect it for simplicity, i.e.,
calculate $J$~(\ref{r2}) at $b\!=\!0$. Considering the case of
very slow proton-spectator with $\bfp_2\sim 0$, we take
$u(p_2)\approx u(0)$ for the IA term $M_a$ in Eqs.~(\ref{r2}). We
also add  the $\pi N$-FSI amplitude $M_c$ with the same
assumptions as for the $N\!N$-FSI, i.e.,
$M_c=-(1/4)\,\sigma^t_{\pi N}\,\langle\cdots\rangle\,J$. Finally,
the FSI correction factor is $R=|M_a\!+\!M_b\!+\!M_c|^2/|M_a|^2$,
and with CD-Bonn DWF~\cite{BonnCD}, we obtain \be
    R=R^{}_{FSI}=\left(\frac{u(0)-0.25\,
    (\sigma^t_{N\!N}\!+\!\sigma^t_{\pi N})\,J}{u(0)}\right)^2
    \!\approx 0.95,
\label{r3}\ee Here, we use some typical values
$\sigma^t_{N\!N}\approx 45$~mb and $\sigma^t_{\pi N}\approx 35$~mb
for the total \crss\ at laboratory momentum $p_{lab}\sim 1 -
1.5$~GeV$/c$. For the integral $J$ at $b\!=\!0$ in Eq.~(\ref{r3})
with CD-Bonn DWF~\cite{BonnCD}, one gets $\dist
J=-(2\pi)^{-1}\!\sum_i c_i\ln m_i$~ in the notations of
Eqs.~(\ref{de2}).

Our Glauber-type calculations are extremely simplified in a number of ways and 
give only a qualitative estimation. Some predictions for the FSI corrections in 
the Glauber approach for $\pi^-$ photoproduction on light nuclei were done in 
Ref.~\cite{HG}. The analysis~\cite{Zhu} of the reaction $\gdpp$ at high energies 
of the photons, based on the approach of Ref.~\cite{HG}, gave the Glauber FSI 
correction of the order of 20\%. Similar values 15\%-30\% for this effect in the 
same approach were obtained in Refs.~\cite{Wei,CLAS}, while our estimation~(\ref{r3}) 
gave smaller value $\sim 5\%$. To comment on this difference in the results, let us 
point out the difference of the approaches used. Here, we use the diagrammatic 
technique. The analyses of Refs.~\cite{Wei,CLAS,HG,Zhu} are based on the approach 
which considers a semi-classical propagation of final particles in the nuclear 
matter. The applicability of the latter approach to the deuteron case is rather 
questionable. Notice that our approximate estimation in terms of Glauber FSI 
correction gives results similar to that obtained with our full dynamical model 
at large angles, i.e., the solid curves in Fig.~\ref{fig:g5}, are in a reasonable 
agreement with the value of $R$ from Eq.~(\ref{r3}).

Thus, we obtain the following behavior of the correction factor
$R$, for the reaction $\gnp$, calculated from the reaction $\gdpp$
at high-energy photon beam with slow proton-spectator. A sizeable
effect $R\ne 1$ is observed in the relatively narrow region
$\theta_1\sim 0$ dominated by the $S$-wave part of NN-FSI with
additional some contribution from the ``Pauli effect" due to the
``suppressed" IA diagram. Small but systematic effect $|R\!-\!1|\ll
1$ is found in the large angular region, where it can be estimated
in the Glauber approach, except for narrow regions close to
$\theta_1\sim 0$ or $\theta_1\sim 180^{\,\circ}$.

\section{Conclusion}
\label{sec:conclusion}

The incoherent pion photoproduction process $\gdpp$ was considered
in a model containing the IA and FSI amplitudes. The $N\!N$- and
$\pi N$-FSI were taken into account. The inputs to the model are
the phenomenological $\gamma N\!\to\!\pi N$, $NN\!\to\!NN$, and
$\pi N\!\to\!\pi N$ amplitudes, the deuteron wave function, and
the additional parameter ($\beta$) for the off-shell behavior of
the ${^1}S_0$ partial amplitude of $pp$-scattering. The
Fermi-motion was also taken into account in the IA amplitudes as
well as in the FSI ($N\!N+\pi N$) terms.

The model reasonably describes the existing data on the \dif\
\crs\ $\dso(\gdpp)$. Sizeable FSI effects were observed at small
laboratory angles $\theta\lesssim 30^{\circ}$ for outgoing pions,
where the main part of the effect comes from the ${^1}S_0$ part of
$pp$-FSI. In this angular range, the theoretical predictions of
different authors reveal the most pronounced disagreements. Thus,
future experiments on the reactions $\gd$ are welcome, especially
at small angles $\theta\lesssim 30^{\circ}$, where data are
absent.

The procedure to extract the \dif\ \crs\ $\dso(\gnp)$ on the
neutron target from the deuteron data was derived in terms of the
FSI correction factor $r$~(\ref{ne3}). To reduce the number of
variables, we gave the results for the averaged correction factor
$R$~(\ref{ne7}), defined as the ratio of the \dif\ \crss\
$\dso_1(\gdpp)$, calculated with full amplitude as well as in the
quasi-free-process approximation, where $\Omega_1$ is the solid
angle of relative motion in the system $\pi^-+$fast~proton. Also
the kinematical cuts with slow spectator proton were used. The
results show a sizeable FSI effect $R\ne 1$, predominantly coming
from the ${^1}S_0$ part of $pp$-FSI, at the angular region close
to $\theta_1\sim 0$, and the region narrows with the increasing
photon energy. In the wide angular range, the effect is small
($|R\!-\!1|\ll 1$) and in agreement with the Glauber estimations.

The more refined analysis requires the use of the factor
$r$~(\ref{ne3}) instead of the averaged one ($R$). Then, we deal
with the ratio of multi-dimensional \dif\ \crss\
$d\sgd^i/d\bfp_2d\Omega_1$, used in Eqs.~(\ref{ne3}). Further, one
should integrate $d\sgd^i/d\bfp_2d\Omega_1$ over the azimuthal
angle $\varphi_1$ in the $\pi^- p_1$ pair, since the \dif\ \crs\
on the neutron in the unpolarized case has no azimuthal
dependence; thus, the \crss\ $d\sgd^i/d\bfp_2d\Omega_1$ turns out
to be a function of 3 variables, \emph{i.e.}, $p_2$, $\theta_2$,
and $\theta_1$ (or $\cos\theta_2$ and $\cos\theta_1$). Thus,
applying Eqs.~(\ref{ne3}) to extract the \dif\ \crs\
$d\sgn/d\Omega_1$ on the neutron, one needs data on the deuteron
\crs\ $d\sgd/d\bfp_2d\Omega_1$ binned in the variables $p_2$,
$\theta_2$, and $\theta_1$, \emph{i.e.}, in the 3-dimensional form.
We plan to discuss this question in detail in the next
publication.



\acknowledgments
\vspace{-3mm}

The authors are thankful to R.~Arndt, W.~Chen, E.~Pasyuk, and
N.~Pivnyuk for useful remarks and interest to the paper. This work was 
supported in part by the U.S.~Department of Energy Grants 
DE--FG02--99ER41110 and DE-FG02--03ER41231,  by the Italian Istituto 
Nazionale di Fisica Nucleare, by the Russian RFBR Grant No.~02--02--16465, 
by the Russian Atomic Energy Corporation ``Rosatom" and by the grant 
NSh-4172.2010.2. V.T. acknowledges The George Washington University Center 
for Nuclear Studies, Jefferson Science Associates, Jefferson Lab, and Dr. 
P.~Rossi for their partial support.

\section{Appendix}
\vspace{-3mm}

\subsection{Invariant amplitudes and phase space}
\label{app:amp}
\vspace{-3mm}

We use standard definitions and the \crs\ of the process
$\,a\!+\!b\!\to\!1+\!\cdots\!+\!n\,$ reads
\be
    \sigma_n=I_n J^{-1}\int\! |M|^2 d\tau_n,
    ~~~~
    d\tau_n=(2\pi)^4\delta^{(4)}(P_i-P_f)
    \prod^n_{i=1} \frac{d^3 \bfp_i}{(2\pi)^3 2E^{}_i}.
\label{amp1}\ee
Here, $M$ is the invariant amplitude; $\,d\tau_n$ is the element of the
final $n$-particle phase space; $P_i(P_f)$ is the total initial (final)
4-momentum; $E^{}_i\,$ and $\bfp_i$ are the total energy and 3-momentum
of the $i$-th final particle; $J\!=4E_am_b\!=4q_{ab}\sqrt{s}\,$ is the
flux factor, where $E_a$ ($m_b$) is the total laboratory energy (mass) of the
particle $a(b)$, $\,q_{ab}$ is the initial relative momentum and
$\sqrt{s}$ is the total CM energy; $I_n\equiv 1/n_1!\cdots n_k!$ is the
identity factor, where $n_i$ is the number of particles of the $i$-th
type ($n_1\!+\!\cdots\!+\!n_k\!=\!n$).


\subsection{Deuteron vertex and wave function}
\label{app:deut}
\vspace{-3mm}

The deuteron vertex $\hgd$, used in Eq.~(\ref{ma1}), can be written
in the form
\be\ba{c}\dist
    \hgd(p)=\frac{g_1}{2m^2}(\epsilon p)+\frac{g_2}{m}\not\!\epsilon,
    \\ \rppt \dist
    g_1=-\frac{3m^2}{p^2}\sqrt{m}\,(p^2\!+\!\alpha^2)w(p),~~
    g_2=\sqrt{m}\,(p^2\!+\!\alpha^2)[\sqrt{2}\,u(p)\!+\!w(p)].
\ea\label{de}\ee
Here, $\epsilon$ is the deuteron polarization
4-vector; $p=|\bfp|$ the relative 3-momentum of the nucleons;
$u(p)$ and $w(p)$ are $S$- and $D$-wave parts of the deuteron wave
function, respectively; $\alpha^2=m\epsilon_d$, where $\epsilon_d$
is the deuteron binding energy. The DWF in $\bfp$-representation
reads \be\ba{c}
    \langle m_1,\tau_1,m_2,\tau_2 |\,\hPsi(\bfp)|\,m_d\rangle
    =\varphi^+_1\,\hPsi(\bfp)\,\varphi^c_2,
    \\ \rppt \dist
    \hPsi(\bfp)=u(p)\hat S_u +w(p)\hat S_w,~~
    \hat S_u=\frac{(\bfsig\bfeps)}{\sqrt{2}},~~
    \hat S_w=\half [(\bfsig\bfeps)-3(\bfn\bfeps)(\bfsig\bfn)].
\ea\label{de1}\ee
Here, $\bfn=\bfp/p$; $\bfeps$ is the deuteron
polarization 3-vector for a given spin state $m_d$; $\,m_i$ and
$\tau_i$ are spin and isospin states of the $i$-th nucleon, and
$\,\varphi^{}_i$ is its spinor and isospinor;
$\varphi^c_i=\tau_2\sigma_2\varphi^*_i$, where $\sigma_2$ and
$\tau_2$ are spin and isospin Pauli matrices. We use the
normalization
$$
\half\int\!\!d\bfp\!\!\sum_{m_1,\tau_1,\,m_2,\tau_2}\!\!\!
|\langle m_1,\tau_1,m_2,\tau_2 |\,\hPsi(\bfp)|\,m_d\rangle|^2
=\int\!\!d\bfp\,\,[u^2(p)+w^2(p)]=(2\pi)^3.
$$
For the DWF of the CD-Bonn potential, the functions $u(p)$ and
$w(p)$ were parameterized~\cite{BonnCD} in the form \be
    u(p)=\!\sum_i \frac{c_i}{p^2\!+\!m^2_i},~~
    w(p)=\!\sum_i \frac{d_i}{p^2\!+\!m^2_i},~~~
    \sum_i\! c_i=\!\sum_i\! d_i=\!\sum_i\! d_i m^2_i
    =\!\sum_i\!\frac{d_i}{m^2_i}=\!0.
\label{de2}\ee
The parameters $c_i$, $d_i$, and $m_i$ are given in the Tables~11
(full model) and 13 (energy-independent model) of Ref.~\cite{BonnCD}.

\subsection{Invariant $\gn$ amplitudes}
\label{app:gn}
\vspace{-3mm}

The general expression for the $\gn$ amplitude $\mgn$ can be
written as \be
    \mgn=\bar u(p_2)\hmgn u(p_1),~~~
    \hmgn=i\sum^4_{i=1}A_i\gamma_5\Gamma_i,
\label{gn1}\ee
where $u(p^{}_{1,2})$ are the nucleon Dirac spinors ($\bar u u=2m$),
$A_i$ are the invariant amplitudes, $\Gamma_i$ are the $4\times 4$
matrices. $\Gamma_i$'s can be taken in the form
\be\begin{array}{ll}
    \Gamma_1=q\sss e\sss, & \Gamma_2=(ep)(qk)-(pq)(ek), \\ \rpt
    \Gamma_3=q\sss (ek)-e\sss (qk),~~~ & \Gamma_4=q\sss (ep)-e\sss (pq),
    ~~~ p=p_1+p_2.
\ea\label{gN2}\ee
Here, $e$ is the photon polarization 4-vector;
$q,k$, and $p_{1,2}$ are 4-momenta of the photon, pion, and
nucleons, respectively. One can write the amplitude
$\mgn$~(\ref{gn1}) in CM frame as \be
    \mgn=8\pi W\,\varphi^+_2\hff\varphi_1,~~~
    \hff=i\he F_1+\hn_2 [\bfsig(\bfn_1\times\bfe)]F_2
    +i\hn_1 (\bfn_2\bfe)F_3 +i\hn_2 (\bfn_2\bfe)F_4 .
\label{gN3}\ee
Here, $\bfe$ is the photon polarization 3-vector;
$\bfq^*$($\bfk^*$) are the photon (pion) CM 3-momenta; $W$ is the
total CM energy; $F_i=F_i(W,z)$ are the CGLN~\cite{CGLN}
amplitudes, $z\!=\cos\theta$; $\varphi^{}_i$ are the Pauli
spinors; $\bfn_1\!=\bfq^*\!/q^*$, $\bfn_2\!=\bfk^*\!/k^*$,
$q^*\!=|\bfq^*|$, $k^*\!=|\bfk^*|$; ``hat" means the product with
$\bfsig$, i.e., $\he=(\bfsig\bfe)$, etc. For unpolarized nucleons
$\dso(\gn)={\dist \frac{k^*}{2q^*}}\,Tr\{\hff\hff^+\}$.
%
%
Equating Eqs.~(\ref{gn1}) with Eqs.~(\ref{gN3}), one finds the relations
between $A_i$'s and $F_i$'s, i.e.,
\be\ba{lll}
    \dist A_1 = \frac{\tff_1+\tff_2}{2W}, &
    \dist ~~\tff_1=\frac{8\pi W}{N_1 N_2}\,F_1, &
    \dist ~~\tff_2=\frac{8\pi W N_1 N_2}{|\bfq||\bfk|}\,F_2,
    \\ \rppt
    \dist A_2 = \frac{\tff_3-\tff_4}{2W}, &
    \dist ~~\tff_3=\frac{8\pi W N_1}{|\bfq||\bfk|W_{\!+}\, N_2}\,F_3, &
    \dist ~~\tff_4=\frac{8\pi W N_2}{|\bfk|^2 W_{\!-}\, N_1}\,F_4,
    \\ \rpptt
    \dist A_3=A_4+A_{34}, &
    \dist ~~A_4=\frac{\tff_2\!+W_{\!+}\,A_1\!-(kq)A_{34}}{W_{\!+}\,W_-}, &
    \dist ~~A_{34}=\frac{W_{\!+}\,\tff_3+W_{\!-}\,\tff_4}{2W},
\ea\label{gN4}\ee
where $W_{\pm}\!=W\pm m$, $N_{1,2}\!=\sqrt{E_{1,2}\!+m}$, and
$E_{1,2}$ are total CM energies of the nucleons.

The isospin structure of the amplitudes $A_i(\gamma N\!\to\!\pi_a N)$
and contributions to the different charge channels read
\be
    A_i =A^{(+)}_i\delta_{a3}\,+\,A^{(-)}_i\half
    [\tau_a,\tau_3]\,+\,A^{(0)}_i\tau_{a\,},
\label{gN5}
\ee
$$
\begin{array}{ll}
    A_i(\gamma p\!\to\!\pi^0 p)=A^{(+)}_i+A^{(0)}_i, &
    ~~A_i(\gamma p\!\to\!\pi^+ n)=\sqrt{2}(A^{(0)}_i+A^{(-)}_i),
    \\ \rpt
    A_i(\gamma n\!\to\!\pi^0 n)=A^{(+)}_i-A^{(0)}_i, &
    ~~A_i(\gamma n\!\to\!\pi^- p)=\sqrt{2}(A^{(0)}_i-A^{(-)}_i).
\end{array}
$$
The amplitudes $A_i(\gn)$ can be obtained from the CGLN~\cite{CGLN}
amplitudes $F_i(\gn)$ through Eqs.~(\ref{gN4}). We use the GW pion
photoproduction amplitudes $F_i$~\cite{pr_PWA}.

\subsection{Matrix elements for $\gn$}
\label{app:agn}

The matrix element $\langle m_2|\,\hmgne|\,\lambda, m_1\rangle$ in
an arbitrary frame can be written in the form \be
    \langle m_2|\,\hmgne|\,\lambda, m_1\rangle=
    N_1 N_2 \langle m_2|L+i(\bfkk\bfsig)|\,m_1\rangle~~~
    (N_i=\sqrt{E_i\!+m}).
\label{agn1}\ee
Making use of Eqs.~(\ref{gn1})-(\ref{gN2}), we obtain
\be\ba{l}
    L =A_1(\bfe[\bfq\times\!(\bfx_1\!-\!\bfx_2)])
    -(\bfe\bfr)(\bfq[\bfx_1\!\times\bfx_2])
    +[A_1 q_0-(qr)](\bfe[\bfx_1\!\times\bfx_2]),
\\
    \bfkk\!=[A_1 c_1+(qr)c_3]\bfe +(\bfe\bfss)\bfq
    +(\bfe\bfss_2)\bfx_1+(\bfe\bfss_1)\bfx_2+A_2 c_2(\bfx_2\!-\bfx_1),
\\
    \bfx_{1,2}=\bfp_{1,2}/(E_{1,2}\!+m),\\ 
    \bfss =A_1(\bfx_1\!+\!\bfx_2)+c_3\bfr,~~
    \bfss_{1,2}=[(qr)-A_1 q_0]\bfx_{1,2}+[(\bfq\bfx_{1,2})-q_0]\bfr,\\
    c_1=q_0(1+\bfx_1\bfx_2),~~ c_2=2[(qp_1)(\bfe\bfk)-(qk)(\bfe\bfp_1)],
    ~~c_3=1-(\bfx_1\bfx_2),\\
    r=A_3 k+A_4 (p_1\!+p_2),~~ \bfr=A_3\bfk+A_4 (\bfp_1\!+\bfp_2).
    \ea
\label{agn2}\ee
Here, $A_i$ are the amplitudes in Eqs.~({\ref{gn1});
$\bfe=\bfe^{(\lambda)}$ is the photon 3-vector, specified by spin
state $\lambda$; $q,k,p_{1,2}$($\bfq,\bfk,\bfp_{1,2}$) are the
4(3)-momenta, defined in Appendix~\ref{app:gn}. We fix two
possible photon states ($\lambda=1,2$) by definition
$e^{(\lambda)}_i\!\!=\delta_{i\lambda}$, where $e^{(\lambda)}_i$
is the $i$-th component of $\bfe^{(\lambda)}$ ($Oz\parallel
\bfq$). Thus, $(\bfe^{(1)}\!\bfe^{(2)})=0$ and
$(\bfe^{(\lambda)}\!\bfq)=0$.

\subsection{Invariant $NN\to NN$ amplitudes}
\label{app:NN}

The NN-scattering matrix depends on 5 independent spin amplitudes,
and different choices can be found in Refs.~\cite{Byst,Hosh}.
In the $N\!N$ rest frame, the $N'_1 N'_2\!\to\!N_1 N_2$ matrix element
can be written in the form
$\langle m^*_1,m^*_2 |\,\hmm_{NN}|\,{m^*_1}',{m^*_2}'\rangle=8\pi W
\langle \hff_{NN}\rangle$, where $W$ is the $N\!N$ effective mass, and
\be
    \langle \hff_{NN}\rangle=\sum^{4}_{i=1}f_i
    (\varphi^+_1\hqq_i \varphi'_1)(\varphi^+_2\hqq_i\varphi'_2)+f_5
    [(\varphi^+_1\hn\varphi^{}_1) (\varphi^+_2\varphi'_2)
    +(\varphi^+_1\varphi^{}_1) (\varphi^+_2\hn\varphi'_2)],
\label{nma1}\ee where $\varphi'_{1,2}$ ($\varphi^{}_{1,2}$) are
the Pauli spinors of the initial (final) nucleons, specified by
spin states $m^{*\prime}_{1,2}$ ($m^*_{1,2}$). Here, we use the
formalism of Ref.~\cite{Hosh}, where $f_1,\cdots f_5$ are the
independent spin amplitudes; $Q_1,\cdots Q_4$ are the $2\times 2$
matrices, and
\be
    Q_1\!=I,~~ Q_2\!=\hn,~~ Q_3\!=\hat m,~~ Q_4\!=\hat l,~~
    \bfn=\!\frac{[{\bfp^*}'\!\times\bfp^*]}{|{\bfp^*}'\!\times\bfp^*|},~~
    \bfm=\!\frac{\bfp^*\!-{\bfp^*}'}{|\bfp^*\!-{\bfp^*}'|},~~
    \bfl=\!\frac{\bfp^*\!+{\bfp^*}'}{|\bfp^*\!+{\bfp^*}'|};
\label{nma2}\ee
${\bfp^*}'\!={\bfp^*_1}'\!-{\bfp^*_2}'$
($\bfp^*\!=\bfp^*_1\!-\bfp^*_2$) is the initial(final) relative
momentum.

To apply Eq.~(\ref{nma1}) for calculation of the matrix elements
$\langle m_1,m_2 |\,\hmm_{NN}|\,m'_1,m'_2\rangle$ in Eq.~(\ref{nn1}),
one should transform the $NN$ amplitude from the deuteron rest frame
to the $NN$ rest frame. The possible way is to transform the nucleon
Dirac spinors to the $NN$ rest frame, and find the corresponding
unitary transformation of spinors in Eq.~(\ref{nma1}), i.e.,
\be
    \varphi\to\huu\varphi,~~~ \huu=N^{-1}(L+i\bfkk\bfsig),~~~
    N=\sqrt{|L|^2+|\bfkk|^2},
\label{nma3}\ee
where $\varphi$ is any of $\varphi^{}_{1,2}$ or $\varphi'_{1,2}$.
The result is
\be\ba{c}
    \dist L=a_0+\bfb\bfx,~~ \bfkk=\bfa+b_0\bfx+[\bfb\times\bfx],~~
    \bfx=\frac{\bfp}{E+m},\\ \rptt
    a_0=c_1 c_2,~~ \bfa=(-s_1 s_2, c_1 s_2, s_1 c_2),~~
    b_0=-x^{}_{N\!N}s_1 c_2,~~
    \bfb=-x^{}_{N\!N}(c_1 s_2, s_1 s_2, c_1 c_2),\\ \rppt\dist
    s_1=\sin\frac{\varphi^{}_{N\!N}}{2},~~
    c_1=\cos\frac{\varphi^{}_{N\!N}}{2},~~
    s_2=\sin\frac{\theta^{}_{N\!N}}{2},~~
    c_2=\sin\frac{\theta^{}_{N\!N}}{2},~~
    x^{}_{N\!N}=\frac{|\bfp|}{E^{}_{N\!N}\!+W}.
    \ea
\label{nma4}\ee
Here: $E$ and $\bfp$ are the total energy and 3-momentum of a given
nucleon in the deuteron rest frame, i.e.,
$\bfp=\bfp^{}_{1,2},\bfp'_{1,2}$ [Fig.~\ref{fig:g1}(b)]; $E^{}_{N\!N}$,
$\bfp^{}_{N\!N}$, $\theta^{}_{N\!N}$, and $\varphi^{}_{N\!N}$ are
the total energy, 3-momentum, polar and azimuthal angles of the
outgoing $N\!N$ system in the deuteron rest frame, respectively.
Finally, for the $N\!N$ matrix elements in Eq.~(\ref{nn1}), we
obtain
\be\ba{c}
    \langle m_1,m_2 |\,\hmm_{NN}|\,m'_1,m'_2\rangle=8\pi W
    \langle \hff_{NN}\rangle, \\ \rptt
    \ba{ll}
    \langle \hff_{NN}\rangle & ={\dist\sum^{4}_{i=1}}f_i\,
    \langle m_1 |\,\huu^+_1\hqq_i\huu'_1|\,m'_1\rangle
    \langle m_2 |\,\huu^+_2\hqq_i\huu'_2|\,m'_2\rangle \\
     & +f_5\,
    [\langle m_1 |\,\huu^+_1\hn\huu'_1|\,m'_1\rangle
    \langle m_2 |\,\huu^+_2\huu'_2|\,m'_2\rangle
    +\langle m_1 |\,\huu^+_1\huu'_1|\,m'_1\rangle
    \langle m_2 |\,\huu^+_2\hn\huu'_2|\,m'_2\rangle].
    \ea\ea
\label{nma5}\ee
One can rewrite the products $\huu\hqq\huu'$ in the form
$\huu\hqq\huu'=V_0+i\bfvv\!\bfsig$, making use of
Eqs.~(\ref{nma2})-(\ref{nma4}), and calculate the factors
$\langle m_i |\cdots |\,m'_i\rangle$ in Eqs.~(\ref{nma5})
(we ommit the details). The Hoshizaki~\cite{Hosh} amplitudes
$f_1,\cdots f_5$ can be expressed through the helicity amplitudes
$H_1,\cdots H_5$ (the relations of $H$'s to other representations
\cite{Byst,Hosh} can be found, for example, in
Ref.~\cite{PWA83}), and we use the results of GW NN partial-wave
analysis~\cite{NN_PWA}.

\subsection{Invariant $\pi N\to\pi N$ amplitudes}
\label{app:piN}

Calculating the $\pi N\!\to\!\pi N$ matrix elements in
arbitrary frame, we start from the invariant amplitude and
write
\be
    M^{}_{\pi N}=\bar u_2(A+B p\sss)u_1=\varphi^+_2\hat\Box\varphi_1.
\label{apin1}\ee
Here, $u_{1,2}(\varphi_{1,2})$ are Dirac (Pauli)
spinors; $A$ and $B$ are the invariant amplitudes;
$p=(p_0,\bfp)=p_{1,2}\!+k_{1,2}$ is the total 4-momentum;
$p_{1,2}\!=(E_{1,2},\bfp_{1,2})$ ($k_{1,2}$) are the 4-momenta of
the initial and final nucleons (pions); $\hat\Box$ is $2\times 2$
matrix. Making use of Eq.~(\ref{apin1}), we obtain \be\ba{c}
    \hat\Box=N(L+i\bfkk\bfsig),~~~ N=\sqrt{(E_1\!+m)(E_2\!+m)},\\
    L=A+Bp_0\!-B(\bfp\,(\bfx_1\!+\!\bfx_2))+(Bp_0\!-A)(\bfx_1\bfx_2),\\
    \bfkk=B[\bfp\times\!(\bfx_2\!-\!\bfx_1)]+(Bp_0\!-A)\,
    [\,\bfx_1\!\times\!\bfx_2],~~ \bfx_{1,2}\!=\bfp_{1,2}/(E_{1,2}\!+m).
\ea\label{apin2}\ee
The matrix elements can be obtained from Eqs.~(\ref{apin2}), i.e.,
$\langle m_2|\hmm^{}_{\pi N}|\,m_1\rangle
=\langle m_2|\hat\Box|\,m_1\rangle$.

In the $\pi N$ rest frame, $\hat\Box=8\pi W
[F\!+iG([\bfn_1\!\times\!\bfn_2]\bfsig)]$, where $F(G)$ is the standard
non-flip (spin-flip) amplitude, $W$ is the effective $\pi N$ mass,
$\bfn_{1,2}\!=\bfp^*_{1,2}/|\bfp^*_{1,2}|$, $\bfp^*_{1,2}$ are the
nucleon CM 3-momenta. Applying Eq.~(\ref{apin1}), one can relate
the amplitudes $A$ and $B$ to $F$ and $G$, i.e.,
\be
    A=4\pi W\left(\frac{F+Gz}{E_+}+\frac{G}{E_-}\right),~~
    B=4\pi\left(\frac{F+Gz}{E_+}-\frac{G}{E_-}\right),~~
    E_{\pm}=E\!\pm m,
\label{apin3}\ee
where $E$ is the nucleon total CM energy, $z$ is the cosine of CM
scattering angle. We use the amplitudes $F$ and $G$, based on the
results of GW $\pi N$ partial-wave analysis~\cite{piN_PWA}.



\end{document}